%% file: main.tex
\documentclass[fleqn,10pt,twocolumn]{wlscirep} 

\usepackage[utf8]{inputenc}
\usepackage[T1]{fontenc}
\usepackage{tablefootnote}
\usepackage{array}
\usepackage[quotient-mode = fraction]{siunitx}
\usepackage{adjustbox}
\usepackage{todonotes}

\usepackage{enumitem} 
\usepackage{algorithm}
\floatname{algorithm}{Procedure}
\usepackage{float}
\usepackage{soul} 
\usepackage{footnote}  
\makesavenoteenv{algorithm} 

\title{Machine-Learning Potentials Predict Orientation- and Mode-Dependent Fracture in Refractory Diborides
}

\author[1,2*]{Shuyao Lin}
\author[3]{Zhuo Chen}
\author[1]{Rebecca Janknecht}
\author[3]{Zaoli Zhang}
\author[2,4]{Lars Hultman}
\author[1]{Paul H. Mayrhofer}
\author[1,2]{Nikola Koutn\'{a}}
\author[2]{Davide G. Sangiovanni}
\affil[1]{Technische Universit\"{a}t Wien, Institute of Materials Science and Technology, Vienna, A-1060, Austria}
\affil[2]{Link\"{o}ping University, Department of Physics, Chemistry, and Biology (IFM), Link\"{o}ping, SE-58183, Sweden}
\affil[3]{Austrian Academy of Sciences, Erich Schmid Institute of Materials Science, Leoben, A-8700, Austria}
\affil[4]{Center for Plasma and Thin Film Technologies, Ming Chi University of Technology, New Taipei City, 24301, Taiwan}

\affil[*]{shuyao.lin@tuwien.ac.at}

\begin{abstract}

\input{0-Abstract}

\end{abstract}

\begin{document}
\flushbottom
\maketitle
\thispagestyle{empty}


\section*{Introduction}

\input{00-Introduction}
\section*{Results and discussion}

\input{1-Vali_Concept}
\input{2-Mode-I}

\input{3-Mix}

\section{Experimental testing and verification of fracture simulations}
\input{3,5-Discussion}

\section{Conclusion}
\input{4-Conclusion}

\section*{Methods}
\input{5-Methods}

\section*{Acknowledgements}
DGS gratefully acknowledges financial support from the Swedish Research Council (VR) through Grant No VR-2021-04426 and the Competence Center Functional Nanoscale Materials (FunMat-II) (Vinnova Grant No. 2022-03071). LH acknowledges financial support from the Swedish Government Strategic Research Area in Materials Science on Functional Materials at Linköping University SFO-Mat-LiU No. 2009 00971. Support from Knut and Alice Wallenberg Foundation Scholar Grants KAW2016.0358 and KAW2019.0290 is also acknowledged by LH.
The financial support (Z.C.and Z.Z.) by the Austrian Science Fund (FWF P33696) is highly acknowledged.
The authors acknowledge original TiB$_{2}$ samples from Anna Viktoria Hirle, also for providing the raw data from a previous publication~\cite{fuger2023tissue} and participating discussion.
PHM and NK acknowledge the Austrian Science Fund, FWF (10.55776/PAT4425523). 
The FFG projects: Bridge and iDAP+ are further acknowledged by PHM.
We sincerely thank the Electron Microscopy Center at USTEM TU Wien for FIB-SEM access.
The computations handling were enabled by resources provided by the National Academic Infrastructure for Supercomputing in Sweden (NAISS) -- at supercomputer centers NSC and PDC, partially funded by the Swedish Research Council through grant agreement no. 2022-06725 -- and by the Vienna Scientific Cluster (VSC) in Austria. 
The authors acknowledge TU Wien Bibliothek for financial support through its Open Access Funding Program.

\section*{CRediT author contributions statement}
{\bf{SL}}:  Methodology, Investigation, Data curation, Formal analysis, Visualization, Writing - original draft. 
{\bf{ZC}}: Investigation, Writing - review \& editing
{\bf{RJ}}: Investigation, Writing - review \& editing
{\bf{ZZ}}, {\bf{LH}}, {\bf{PHM}}: Resources, Funding acquisition, Writing - review \& editing
{\bf{NK}}: Conceptualization, Methodology, Supervision, Writing - review \& editing
{\bf{DGS}}: Conceptualization, Methodology, Supervision, Software, Investigation, Writing - review \& editing

\section*{Conflicts of interest}
The authors declare that they have no known competing financial interests or personal relationships that could have appeared to influence the work reported in this paper.

\section*{Data availability}
Data will be made available on request.

\bibliographystyle{ieeetr}
\bibliography{references}

\end{document}

%% file: 0-Abstract.tex
Fracture toughness ($K_\mathrm{Ic}$) and fracture strength ($\sigma_\mathrm{f}$) are key criteria in the selection and design of reliable ceramics. However, their experimental characterization remains challenging--especially for ceramic thin films, where size and interfacial effects hinder accurate and reproducible measurements. Here, machine-learning interatomic potentials (MLIPs) trained on \textit{ab initio} datasets of single crystal models deformed up to fracture are used to characterize transgranular cleavage in pre-cracked ceramic diboride TMB$_2$ (TM = Ti, Zr, Hf) lattices through stress intensity factor ($K$)-controlled loading. Mode-I simulations performed across distinct crack geometries show that fracture is primarily driven by straight crack extension along the original plane. The corresponding macroscale fracture-initiation properties ($K_\mathrm{Ic} \approx 1.7$--$2.9$~MPa$\cdot\sqrt{\text{m}}$, $\sigma_\mathrm{f} \approx 1.6$--$2.4$~GPa) are extrapolated using established scaling laws. Considering TiB$_2$ as a representative system, additional simulations explore loading conditions ranging from pure Mode-I (opening) to Mode-II (sliding). TiB$_2$ models containing prismatic cracks exhibit their lowest fracture resistance under mixed-mode conditions, where the crack deflects onto pyramidal planes--as confirmed by nanoindentation tests on TiB$_2$(0001) thin films. This study establishes $K$-controlled, MLIP-based simulations as predictive tools for orientation- and mode-dependent fracture in ceramics. The approach is readily extendable to finite temperatures for evaluating fracture behavior under conditions relevant to refractory applications.

%% file: 00-Introduction.tex
 Fracture mechanics roots in Griffith's work during the 1920s~\cite{griffith1921vi}. Griffith recognized that hard yet brittle materials, such as glass, fracture at stresses far below their theoretical strength due to microstructural imperfections. These flaws act as stress concentrators that facilitate crack initiation. He also demonstrated that crack propagation occurs when the elastic strain energy released during crack extension exceeds the energy required to form new crack surfaces. 
 
 Three decades later, Irwin~\cite{irwin1957analysis} extended Griffith’s energy-based criterion by incorporating dissipative, non-linear processes such as plastic deformation in metals. He introduced the concept of stress intensity factor, \( K \), to quantify near-tip stress field in linear elastic materials. Irwin also classified fracture into three distinct modes - Mode I (opening), Mode II (sliding), and Mode III (tearing) - each governed by a corresponding critical stress intensity factor: \(K_\text{Ic}\), \(K_\text{IIc}\), and \(K_\text{IIIc}\).
 
 The critical stress intensity factor \(K_\text{Ic}\) is recognized as the practically most significant. The term \(K_\text{Ic}\) is often referred to as the material's fracture toughness, since fracture in solids generally initiates at structural flaws under tension. The \(K_\text{Ic}\) is a property measured through standardized tests (see, e.g., Refs.~\cite{nose1988evaluation,anderson1985elastic,shaji2003plane,underwood1984review,kolhe1998effects}), which involve introducing a crack into a sample and applying a controlled load or displacement to determine the conditions leading to fracture. However, the reproducibility of  \(K_\text{Ic}\) measurements depends strongly on microstructural features, including crystallographic defect density ~\cite{gludovatz2010fracture,launey2009fracture,zeng2010modeling,moller2014fracture,shimokawa2011roles,samborski2010dynamic,kleebe1999microstructure}.

Traditional fracture toughness tests are limited in resolving nanoscale mechanisms responsible for fracture initiation. These limitations become especially pronounced in thin-film materials, where small dimensions and substrate effects complicate measurements ~\cite{zhang2021correlating}. Thus, advancing the fundamental understanding of fracture requires complementing mechanical tests with state-of-the-art {\textit{in-situ}} imaging ~\cite{ma2021situ,fadenberger2010situ}. In this context, atomistic simulations play a vital role by providing direct insights into fracture initiation mechanisms and enabling accurate characterization of fracture properties at atomic scale.

Machine-learning interatomic potentials (MLIPs) can offer detailed atomic-scale understanding of materials' fracture initiation and accurate evaluation of fracture properties. 
Although the reliability of MLIP-based simulations depends on the quality of the underlying {\textit{ab initio}} training data~\cite{mishin2021machine,dragoni2018achieving,mueller2020machine,zhang2024efficiency,behler2016perspective}, MLIPs are systematically improvable force fields ~\cite{deringer2019machine,zuo2020performance} ~\cite{smith2017ani,shapeev2020elinvar,MLIP} that can achieve accuracy comparable to density functional theory (DFT), but with up to five orders of magnitude greater computational efficiency. 
We have recently proposed an MLIP training workflow and a validation standard~\cite{lin2024machine} aimed at reproducing the elastic and plastic properties of bulk ceramic lattices -- specifically, systems free of extended crystallographic defects in their unstrained state -- as predicted by {\textit{ab initio}} molecular dynamics (AIMD). 
Using MLIP-based molecular dynamics simulations, we investigated how theoretical strength varies with supercell size, as well as the anisotropy of slip and fracture behavior under tensile and shear loading at both room and elevated temperatures. 
The training and validation strategy presented in Refs.~\cite{lin2024machine,lin2024shear,koutna2025machine} lay the foundations to this work. 

Here, we use MLIPs in $K$-controlled-loading simulations to investigate the mechanical properties and fracture paths in defective (pre-cracked) Group-IV transition-metal diborides, TMB$_{2}$:s (TM$=$Ti, Zr, Hf). 
As most ceramics, TMB$_{2}$:s are prone to fracture without yielding, but exhibit ultra-high thermal stability~\cite{sevik2022high}, exceptionally high hardness~\cite{holleck1986material}, corrosion resistance, and excellent thermal and electrical conductivity~\cite{wang1995electrical}. 
Unlike other hard ceramic protective coatings (e.g. nitrides and carbides) for which the fracture properties have been characterized relatively thoroughly~\cite{waldl2022evolution,moritz2021microstructure,daniel2017grain,csanadi2020small,tatami2015local,best2016small} 
, information on the toughness and strength available in the literature for TMB$_{2}$:s is sparse~\cite{monteverde2003advances,csanadi2024effect,vidivs2024hardness}.
An experimental characterization of the fracture resistance of diboride films is further complicated by largely varying degrees of stoichiometry (TMB$_{1.53\text{--}2.72}$) which, together with a typically high density of crystallographic defects~\cite{hu2024influence,glechner2022influence}, constitutes an additional hurdle to understanding the structure/property relationship in this class of materials~\cite{fuger2023tissue}. The limitations of the experiments provide further motivation to use MLIPs to characterize fracture initiation in diboride systems. 

%% file: 1-Vali_Concept.tex
\section{MLIP validation and theoretical strengths of defect-free crystals}

\begin{table*}[!h]
\centering
\small
\caption{\footnotesize 
{\bf{MLIP validation.}} Comparison between ML-MS and DFT results for diboride lattice constants ($a$, $c$), elastic constants ($C_{ij}$), polycrystalline moduli (Young's modulus, $E$, bulk modulus, $B$, shear modulus, $G$), and relaxed surface energies $E_{\text{surf}}$ on (0001), $(10\overline{1}0)$, and $(\overline{1}2\overline{1}0)$ lattice planes. Literature DFT values are also included.}
\resizebox{\textwidth}{!}{
\begin{tabular}{ccccccccccccccccc}
    \hline
    \hline
    TMB$_{2}$ & \multicolumn{2}{c}{Lattice constants (\AA) }& \multicolumn{8}{c}{Elastic constants (GPa)} & \multicolumn{3}{c}{Surface energy (J/m$^{2}$)}&\multicolumn{1}{c}{ Reference }\\
      { } & $a$ & $c$ & $C_{11}$ & $C_{33}$ & $C_{44}$  & $C_{12}$ & $C_{13}$ & $E$ & $B$ & $G$ & $(0001)$ & $(10\overline{1}0)$ & $(\overline{1}2\overline{1}0)$ & { } \\
    \hline
    { } & {3.027}  & {3.213} & {654} & {464} & {259} & {76} & {115} & {580} & {263} & {256} & {3.80} & {3.98} & {3.42} & {This work (DFT)}\\
    { } &  {3.030} & {3.204} & {636} & {441} & {272} & {61} & {92} & {576} & {242} & {261} & {3.80} & {4.12} & {3.57}&{This work (ML-MS)}\\
    {TiB$_{2}$} & {3.029}  & {3.219} & {656} & {461} & {259} & {65} & {98} & {582$^{I}$} & {253$^{I}$} & {261$^{I}$} & {/} & {/} & {/} & {Ref.~\cite{zhou2015general}}\\
    { } & {/}  & {/} & {660} & {464} & {255} & {60} & {96} & {583$^{I}$} & {250} & {258} & {4.20$^{II}$} & {4.10$^{II}$} & {/} & {Ref.~\cite{gan2021robust}}\\
    { } & {3.029}  & {3.219} & {/} & {/} & {/} & {/} & {/} & {/} & {/} & {/} & {4.21} & {4.19} & {/} & {Ref.~\cite{sun2017anisotropic}}\\
    \hline
    { } & {3.180}  & {3.545} & {539} & {422} & {270} & {52} & {109} & {523} & {226} & {235} & {3.86} & {4.54} & {3.83} & {This work (DFT)}\\
    { } & {3.167}  & {3.531} & {551} & {427} & {248} & {54} & {118} & {517} & {234} & {228} & {3.61} & {4.38} & {3.57} & {This work (ML-MS)}\\
    {ZrB$_{2}$} & {3.168}  & {3.536} & {555} & {436} & {254} & {62} & {119} & {524} & {238} & {231} & {/} & {/} & {/} & {Ref.~\cite{zhou2015general}}\\
    { } & {/}  & {/} & {539} & {420} & {238} & {60} & {116} & {502$^{I}$} & {231} & {218} & {3.85$^{II}$} & {4.45$^{II}$} & {/} & {Ref.~\cite{gan2021robust}}\\
    { } & {3.168}  & {3.536} & {/} & {/} & {/} & {/} & {/} & {/} & {/} & {/} & {3.91} & {4.33} & {/} & {Ref.~\cite{sun2016theoretical}}\\
    \hline
    { } & {3.149}  & {3.480} & {602} & {432} & {309} & {57} & {105} & {580} & {238} & {266} & {3.88} & {4.53} & {4.07} & {This work (DFT)}\\
    { } & {3.127}  & {3.473} & {604} & {473} & {271} & {70} & {137} & {565} & {263} & {248} & {3.53} & {4.51} & {3.83} & {This work (ML-MS)}\\
    {HfB$_{2}$} & {3.165}  & {3.512} & {584} & {457} & {257} & {98} & {135} & {544} & {253} & {238} & {/} & {/} & {/} & {Ref.~\cite{zhou2015general}}\\
    { } & {/}  & {/} & {588} & {448} & {248} & {89} & {138} & {533$^{I}$} & {260} & {227} & {3.80$^{II}$} & {4.35$^{II}$} & {/} & {Ref.~\cite{gan2021robust}}\\
    { } & {3.163}  & {3.515} & {602} & {452} & {258} & {78} & {137} & {550} & {260} & {239} & {/} & {/} & {/} & {Ref.~\cite{yang2023first}}\\
\hline
\hline
\end{tabular}
}
\begin{flushleft}
$^{I}$: Calculated based on the reference data.\\
$^{II}$: Value extracted from a figure.\\
\vspace{-0.36cm}
\end{flushleft}
\label{SE}
\end{table*}

Our MLIPs, based on the moment tensor potential (MTP) framework ~\cite{MLIP}, are trained on {\textit{ab initio}} molecular dynamics (AIMD) data for transition metal diborides subjected to deformation up to fracture. All Group-IV diborides (TiB$_2$, ZrB$_2$, HfB$_2$) crystallize in the hexagonal $\alpha$/AlB$_2$-type phase~\cite{hu2025microstructure}.
Training and validation errors, quantified by the residual mean square error (RMSE), remain below 8~meV/atom for energies, 0.24~eV/{\AA} for forces, and 0.6~GPa for stresses, as reported in \textcolor{black}{Tab.~S1 of the Supplementary Material}.
Before applying these MLIPs to simulations of defective diboride crystals under load, their reliability is demonstrated through direct comparison with DFT results and AIMD-based tensile tests.

Tab.~\ref{SE} presents the results obtained by MLIP-based molecular statics simulations and DFT calculations from this and previous studies. 
Although our MLIPs are trained mainly on {\it{ab initio}} MD data collected at finite temperature, they reproduce 0~K DFT lattice parameters ($a$, $c$), elastic constants ($C_{ij}$), and surface energies ($E_{\text{surf}}$) satisfactorily well. 
Note that DFT and AIMD calculations of this work are carried out using the same accuracy parameters (see the {\bf{Methods}}). 
The deviations in lattice parameters are below 0.7\%, while the $C_{11}$, $C_{33}$, and $C_{44}$ elastic constants, as well as the polycrystalline bulk ($B$), shear ($G$) and Young’s moduli ($E$) differ by less than 10\% from the corresponding DFT values. 
For $C_{i\neq j}$ elastic constants, the deviations are generally below 20\%.
With regard to surface energies, the differences between MLIP and DFT values are lower than 6.5\%. 

\begin{figure}[!h]
\centering
\includegraphics[width=0.45\textwidth]{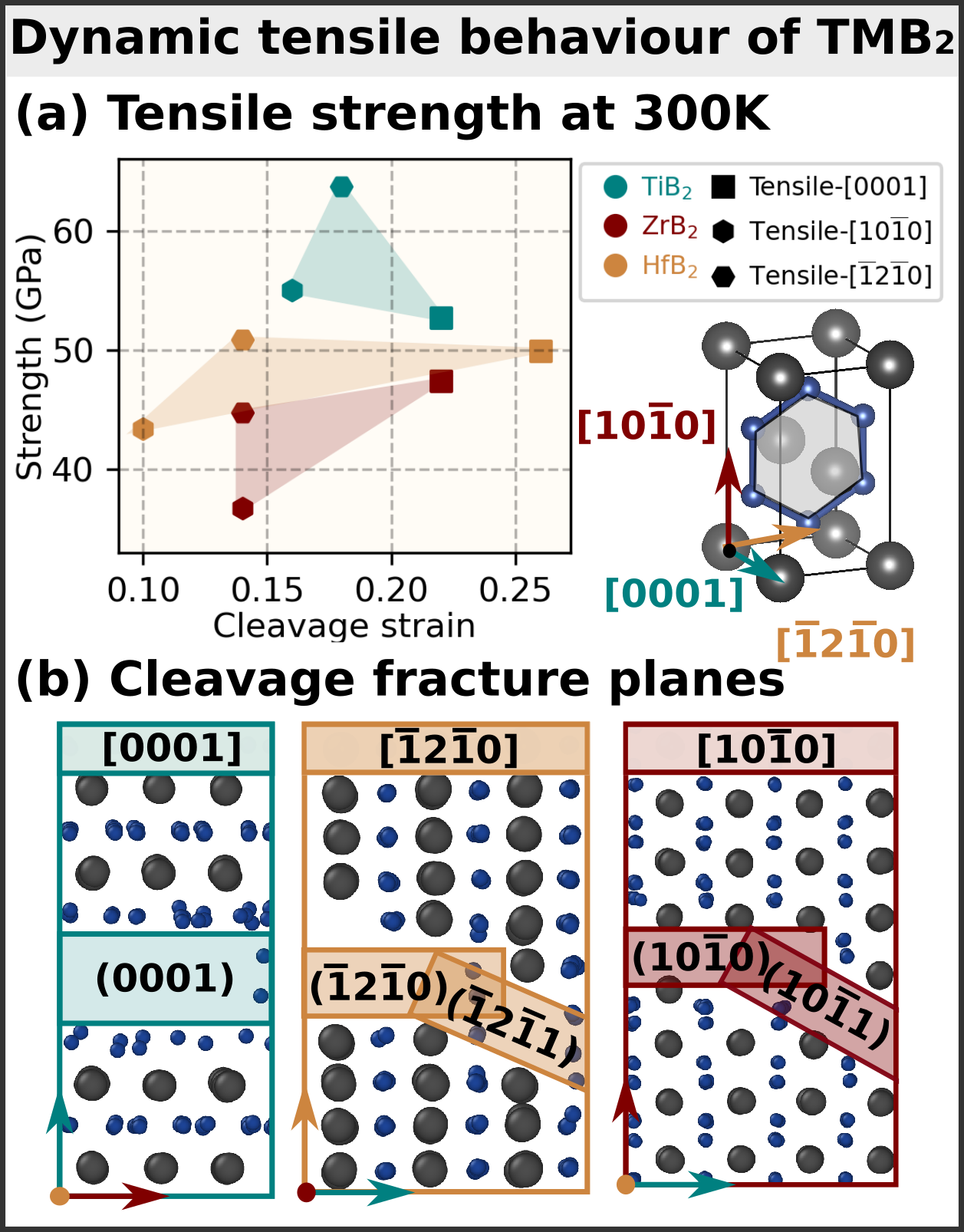}
\caption{\footnotesize 
{\bf{Uniaxial tensile simulations at 300~K using MLIP-based molecular dynamics (ML-MD).}}
(a) Maximum stress (theoretical tensile strength) sustained by TMB$_{2}$:s, (TM$=$Ti, Zr, Hf) during uniaxial tensile deformation along [0001] (square), $[10\overline{1}0]$ (vertical hexagon), and $[\overline{1}2\overline{1}0]$ (horizontal hexagon) directions.
(b) Cleavage fracture after reaching the maximum stress point, with plane identification. 
The image on the right in (a) shows a diboride hexagonal lattice structure in the $\alpha$-phase, including orthogonal crystal axes. 
The shadowed boron hexagonal layer is aligned with the basal plane.
}
\label{Vali}
\end{figure}

In addition to reproducing elastic properties, our MLIPs describe materials' behaviors under stress conditions that induce structural instabilities. 
Specifically, the potentials were trained on AIMD data of supercells subjected to uniaxial tension up to cleavage and to shear strain up to activation of lattice slip at finite temperatures.
Thus, the training sets encompass diborides under tensile strain along various crystallographic directions and shear deformation along distinct slip systems.
Training and extensive validation of the TiB$_2$-MLIP is detailed in Refs.~\cite{lin2024machine,lin2024shear}. 
Importantly, the ability of the MLIP to model tensile and shear deformation in large single-crystal supercells -- beyond the feasible size limits of DFT calculations -- has been assessed using the concept of the extrapolation grade ($\gamma$~\cite{MV}), which is commonly employed in active learning to identify {\it{extrapolative}} atomic environments. More precisely, $\gamma$ quantifies the similarity between local atomic environments that emerge during nanoscale simulations and those represented in the training set. A high extrapolation grade indicates significant deviation from the training data and corresponds to greater uncertainty in the predicted energies and forces. 

Since the newly trained MLIPs for HfB$_2$ and ZrB$_2$ follow the same training protocol previously established for TiB$_2$,~\cite{lin2024machine,lin2024shear} we do not repeat all validation steps here (see \textcolor{black}{the Supplementary Material with Fig.~S2}). Overall, the results obtained using our force fields closely reproduce the properties of the underlying AIMD training configurations. These include cleavage mechanisms on various low-index planes, which are particularly relevant for the crack propagation simulations presented in this study.   

We begin by analyzing the results of MLIP-based molecular dynamics (MLIP-MD) simulations performed on small ( $\sim10^3$ atoms), initially defect-free, single-crystal diboride supercells subjected to tensile strain at room temperature. While these simulations are primarily intended to capture temperature-dependent elastic responses (e.g., Ref.~\cite{sangiovanni2021temperature}) and ideal fracture properties -- such as the theoretical strength and intrinsic toughness of a perfect crystal (e.g., Ref.~\cite{sangiovanni2023valence}) -- they also allow for the rapid identification and clear visualization of energetically preferred fracture planes. In addition, they can signal whether the crystal has an inherent tendency to undergo stress-induced lattice transformations, which may enhance toughness and delay fracture initiation.~\cite{sangiovanni2023valence,koutna2022atomistic}. As shown in our earlier studies on TiB$_{2}$~\cite{lin2024machine}, the fracture mechanisms observed under uniform tensile strain of pristine lattices remain qualitatively unchanged with increasing supercell size. This supports the use of small, computationally efficient models to explore the intrinsic fracture response of ideal crystals in different loading orientations.

Fig.~\ref{Vali} summarizes the theoretical tensile strength and the corresponding fracture strain of diborides elongated parallel to [0001], [10$\overline{1}$0], [$\overline{1}$2$\overline{1}$0] crystallographic directions. The results reveal a pronounced anisotropy in the mechanical response: while the {\it{elastic}} response of hexagonal lattices is isotropic within the basal plane, the tensile strength and fracture strain vary significantly with loading orientation. Among the studied materials, TiB$_{2}$ consistently exhibits the highest theoretical strength, followed by HfB$_{2}$ and ZrB$_{2}$ (Fig.~\ref{Vali}-(b)).
All systems withstand the largest strain along the [0001] axis.

As shown in Fig.~\ref{Vali}-(b), fracture occurs through brittle cleavage along basal, prismatic, or pyramidal planes, depending on the direction of applied tension. While the active cleavage plane changes with orientation, the fracture mechanisms remain consistent across the three diborides. Interestingly, in some cases the fracture planes are not orthogonal to the loading axis, but instead align with first-order pyramidal planes such as (\({10\bar{1}1}\)). This result is somewhat unexpected, but could be explained based on energetic arguments (differences in surface energies) or shear stress accumulation during tension. 

Fracture along first-order pyramidal planes of hexagonal crystals has been predicted in theoretical studies based on surface energy, interplanar spacing, and elastic modulus. For example, previous analyses identified these planes as energetically favorable cleavage paths under specific stress orientations -- particularly in hcp titanium and magnesium~\cite{Paidar2020}. 
Experimental studies further support this view, having observed crack propagation along (\({10\bar{1}1}\))-type planes in single-crystal Ti and Mg~\cite{Kaushik2014,Mine1998}. Our MLIP-MD simulations are consistent with these findings: in addition to fracture along low-index cleavage planes, we observe skewed crack paths that align with first-order pyramidal planes. This suggests that crystallographic anisotropy in AlB$_{2}$-type diborides can promote cleavage along inclined planes, even under nominally uniaxial tension. 

Although these results provide valuable insight into the intrinsic fracture behavior of ideal, defect-free crystals, they should be interpreted with caution. The small size of the simulation cells and the absence of native extended defects -- common in sputter-deposited diborides -- limit the reliability of the predicted crack paths. A more physically representative description of fracture initiation is offered by nanoscale simulations based on $K$-controlled loading of defective lattices (see below).

%% file: 2-Mode-I.tex
\section{Mode-I loading of TMB$_{2}$:s}

\begin{figure*}[!h]
\centering
\includegraphics[width=1.0\textwidth]{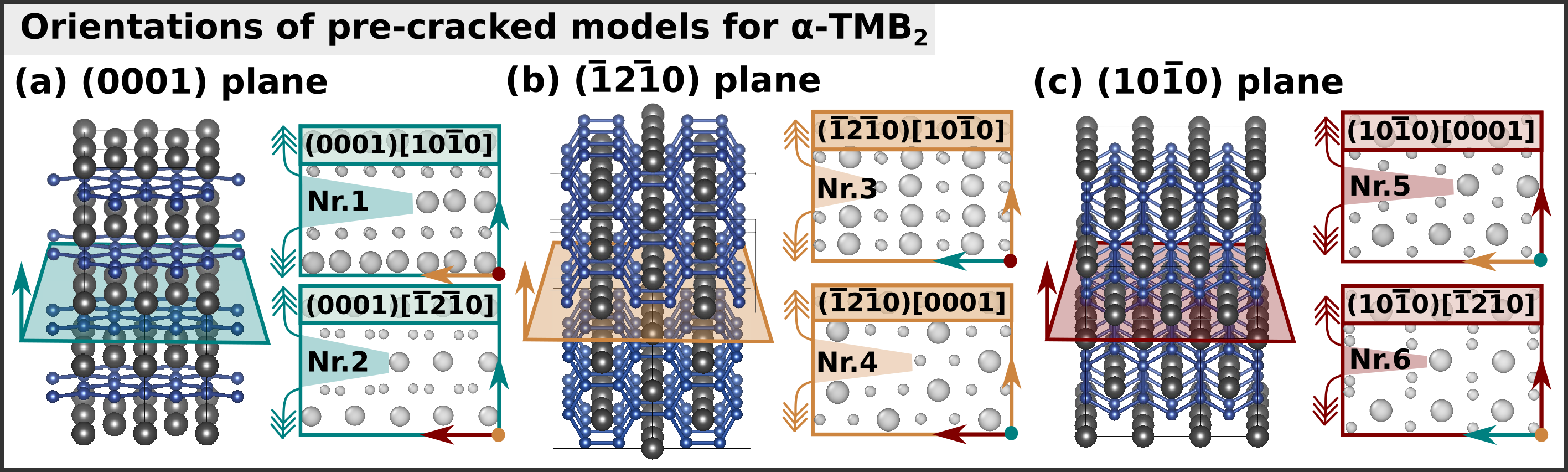}
\caption{\footnotesize 
{\bf{Cracked-plate lattice geometries of  hexagonal $\alpha$-structured diborides considered in this work.}} (a) (0001) crack surface with $[10\overline{1}0]$ (Nr.1) and $[\overline{1}2\overline{1}0]$ (Nr.2) crack-front directions. 
(b) $(\overline{1}2\overline{1}0)$ crack surface with $[10\overline{1}0]$ (Nr.3) and [0001] (Nr.4) crack-front directions. 
(c) $(10\overline{1}0)$ crack surface with [0001] (Nr.5) and $[\overline{1}2\overline{1}0]$ (Nr.6) crack-front directions.
}
\label{Geo}
\end{figure*}
The mechanical properties of diborides are calculated via \(K\)-controlled loading (molecular statics at 0~K) of lattice models containing an atomically sharp crack (cracked-plate models). In these simulations, the stress-intensity factor \(K_I\) is incrementally increased to identify the conditions corresponding to the onset of crack extension~\cite{andric2018atomistic,sangiovanni2024controlled}. Calculations are repeated for varying plate areas \(A\), enabling extrapolation of the fracture-initiation toughness and fracture-initiation strength to the macroscale limit using established scaling relations~\cite{sangiovanni2024controlled}. Here, both properties are extracted from the tensile stress versus stress-intensity curve at the value of \(K_I\) corresponding to the rupture of the first chemical bond at the crack tip.

To systematically investigate fracture behavior in hexagonal \(\alpha\)-structured TMB\(_2\), we consider six low-index crack configurations, denoted as \((hklm)[h'k'l'm']\), where the first set indicates the crack plane and the second the crack-front direction (Fig.~\ref{Geo}). Among these, the $(\overline{1}2\overline{1}0)[10\overline{1}0]$ and $(10\overline{1}0)[\overline{1}2\overline{1}0]$ orientations (Nr.3 and Nr.6), in which fracture initiates perpendicular to (0001) basal plane, are particularly relevant for comparison with experiments. This is because the [0001] direction corresponds to the typical growth orientation of TMB\(_2\) thin films, making these geometries representative of loading conditions encountered in experimental toughness assessments (e.g., by nanoindentation).

\begin{figure*}[!h]
\centering
\includegraphics[width=1.0\textwidth]{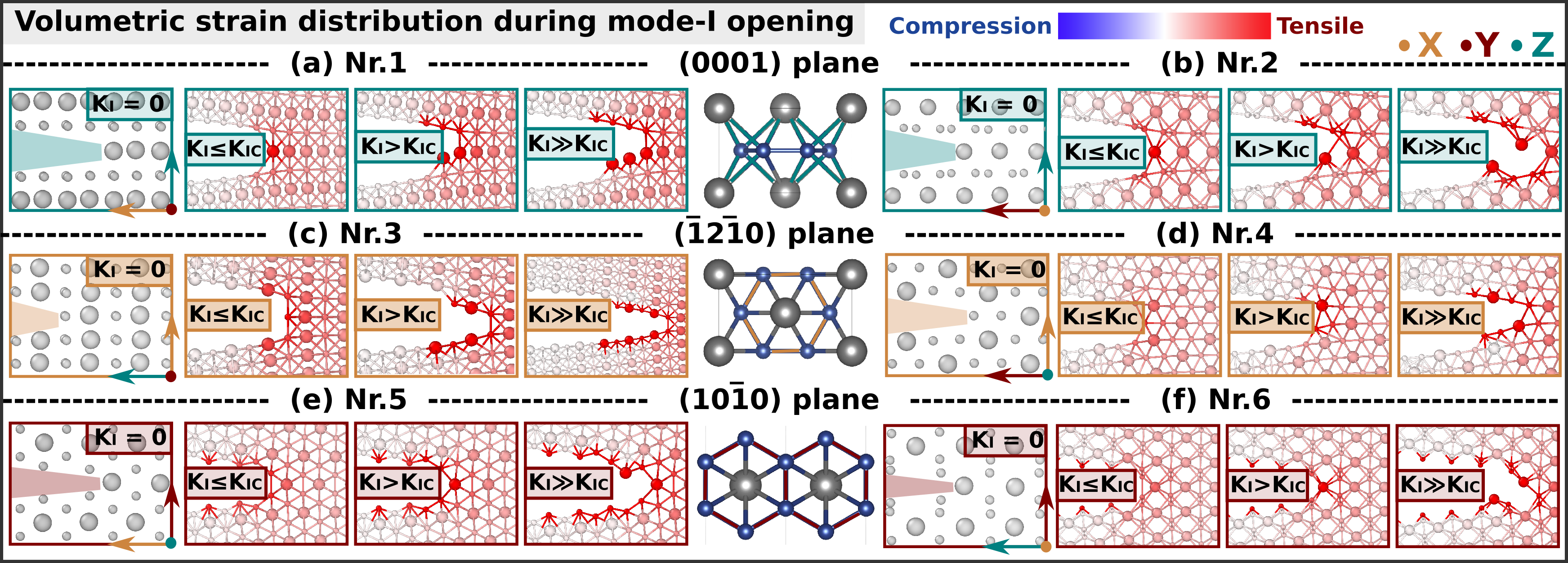}
\caption{\footnotesize 
{\bf{Bond breakage and volumetric strain distribution in cracked plate models subjected to Mode-I loading as a function of the stress intensity factor.}} Atomic configurations just before, shortly after, and well above $K_{Ic}$ for different crack geometries: (a, b) (0001), (c, d) $(\overline{1}2\overline{1}0)$, and (e, f) $(10\overline{1}0)$ planes. 
The simulation snapshots illustrate volumetric strain distributions in blue/red color scale. 
}
\label{Bond}
\end{figure*}

Fig.~\ref{Bond} illustrates the evolution of volumetric strain and bond-breaking events during Mode-I loading for each crack geometry, showing atomic configurations at increasing \(K_I\): before crack extension, at the onset of propagation, and well beyond it. In all cases, stress concentrates at the crack tip as \(K_I\) approaches the critical value \(K_{Ic}\), triggering fracture. On the (0001) plane (Fig.~\ref{Bond}a,b), cracks initiate via rupture of TM--B bonds, while on the \((1\bar{2}10)\) and \((10\bar{1}0)\) planes (Fig.~\ref{Bond}c-f), fracture begins with B--B bond breaking. Most commonly, fracture initiation is marked by the breaking of a single bond, although multiple bond ruptures are observed in some cases, as in model Nr.3. In certain configurations, crack advance is delayed by \textit{crack trapping} -- a phenomenon arising from the discreteness of atomic bonding, which can locally stabilize the crack tip over a finite loading interval before mechanical instability sets in~\cite{Thomson1971}.

Following crack initiation, the crack propagation behavior varies by geometry. Some cracks extend smoothly along the original cleavage plane (e.g., Nr.1, Nr.3, Nr.4), while others follow more complex trajectories, such as zigzag growth (Nr.2, Nr.5) or oblique deflection across crystallographic planes (Nr.6). These patterns underscore the importance of lattice anisotropy and local bonding topology in guiding fracture evolution, as discussed in detail below.

\begin{figure*}[!h]
\centering
\includegraphics[width=1.0\textwidth]{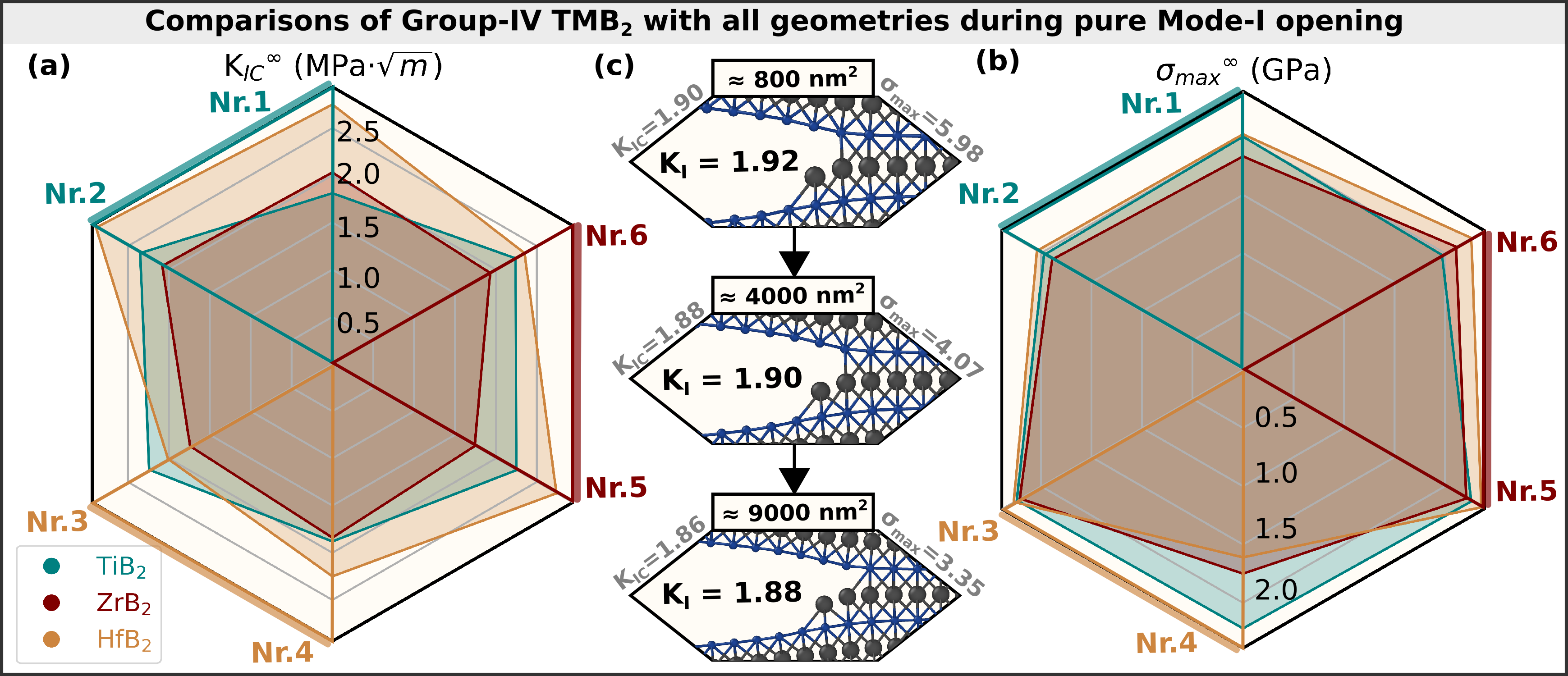}
\caption{Comparison of extrapolated Mode-I fracture toughness \(K_{Ic}^{\infty}\) (a) and fracture strength \(\sigma_{\text{max}}^{\infty}\) (b) for TiB\(_2\) (teal), ZrB\(_2\) (orange), and HfB\(_2\) (red), across all six crack geometries. The geometry indices \(Nr.1\)--\(Nr.6\) are also colored in teal--orange--red, consistent with Fig.~\ref{Geo}. Note that all crack geometries are modeled for all three materials.
(c) Illustration of the influence of the plate area on the calculated fracture toughness (case of TiB\(_2\) with crack geometry \(Nr.1\)). Note that the fracture mechanism remains qualitatively unchanged with increasing size of the supercell.}
\label{mode1comp}
\end{figure*}

Fig.~\ref{mode1comp} compares the extrapolated fracture toughness \(K_{Ic}^{\infty}\) (Fig.~\ref{mode1comp}a) and maximum fracture strength \(\sigma_{\text{max}}^{\infty}\) (Fig.~\ref{mode1comp}b) for Group-IV TMB\(_2\) systems across all six geometries. These values are obtained through inverse polynomial extrapolation to infinite plate area, following the procedure in Ref.~\cite{sangiovanni2024controlled}, with additional details provided in the Supplementary Material (Fig. S3). HfB\(_2\) generally exhibits the highest \(K_{Ic}^{\infty}\), near 2.8~MPa$\cdot\sqrt{m}$, except in geometry Nr.3, where TiB\(_2\) shows a higher toughness (2.24 vs. 2.01~MPa$\cdot\sqrt{m}$). ZrB\(_2\) typically shows the lowest values (1.8~MPa$\cdot\sqrt{m}$), except in Nr.1, where it marginally exceeds TiB\(_2\). Maximum strengths \(\sigma_{\text{max}}^{\infty}\) are comparable across materials (2.0~GPa), with the exception of geometry Nr.4, where TiB\(_2\) sustains higher stress prior to fracturing. Importantly, while toughness and strength vary with plate size -- a fundamental reason for performing \(K\)-controlled simulations, which require sufficiently large plate areas to avoid biasing crack-tip phenomena -- the underlying fracture mechanisms remain qualitatively unchanged for a given crack geometry across all materials (Fig.~\ref{mode1comp}c). This size invariance supports the validity of using the fracture properties of finite-size models to extrapolate macroscale fracture properties.

It is instructive to compare the extrapolated toughness values obtained from atomistic simulations (\(K_{Ic}^{\infty}\)) with the corresponding Griffith-model estimates (\(K_{Ic}^{G}\)), as the ratio \(K_{Ic}^{\infty} / K_{Ic}^{G}\) provides a meaningful metric of intrinsic brittleness or plasticity-mediated toughening. Griffith predictions are derived from unrelaxed surface energies and zero-Kelvin elastic constants (Table~\ref{SE})~\cite{andric2019atomistic,sangiovanni2023descriptor,sangiovanni2024controlled}. A ratio near unity reflects ideally brittle behavior, while significantly higher values indicate enhanced toughness due to plastic deformation at the crack tip. For instance, ratios approaching three have been reported for high-Al-content Ti$_{1-x}$Al$_x$N~\cite{sangiovanni2024controlled}, where atomistic simulations reveal crack-tip plasticity. In contrast, the near one-to-one correspondence observed here for TMB\(_2\) (Tab.~\ref{Comp}) reflects their intrinsically brittle nature, consistent with other brittle ceramics such as TiN(001), where \(K_{Ic}^{\infty} \approx K_{Ic}^{G}\) and no plasticity is observed for atomically sharp cracks~\cite{sangiovanni2024controlled}. Nevertheless, deviations between \(K_{Ic}^{\infty}\) and \(K_{Ic}^{G}\) -- typically within 0.5--20\%, but reaching up to 30\% for HfB\(_2\) geometries Nr.1 and Nr.2 -- underscore the limitations of the Griffith model. Such differences may arise not only from atomistic-scale phenomena like lattice trapping, bond discreteness, and local stress localization~\cite{sangiovanni2023descriptor}, but also from crack path deflection, as observed in TiN(111) reorienting onto lower-energy (001) planes (Fig.~8a in Ref.~\cite{sangiovanni2024controlled}). Unlike Griffith's assumption of linear elasticity and straight-through propagation along the initial cleavage plane, \(K\)-controlled atomistic simulations resolve these non-ideal behaviors directly, making them essential for accurate predictions of fracture in brittle ceramics.

\begin{table*}[!h]
\centering
\small
\caption{\footnotesize 
{\bf{Macroscale Mode-I fracture initiation toughness ($K_{Ic}^{\infty}$), Griffith fracture toughness ($K_{Ic}^{G}$), and fracture strengths ($\sigma_{max}^{\infty}$).}} 
The $K_{Ic}^{\infty}$, $\sigma_{max}^{\infty}$, and their standard deviations are extrapolated at the infinite size limit by fitting the $K_{Ic}$ and  $\sigma_{max}$ values calculated for finite plate areas (see constitutive scaling laws established in Ref.~\cite{sangiovanni2024controlled}). The $K_{Ic}^{G}$ is derived from the unrelaxed surface energy ($E_{surf}^{unrel}$) and elastic tensor computed by ML-MS.}
\resizebox{\textwidth}{!}{
\begin{tabular}{ccccccccccccccccc}
    \hline
    \hline
    \footnotesize 
    Geometry & \multicolumn{3}{c}{TiB$_{2}$ }& \multicolumn{3}{c}{ZrB$_{2}$} & \multicolumn{3}{c}{HfB$_{2}$}\\
      { } & K$_{Ic}^{\infty}$ (MPa$\cdot\sqrt{m}$) & K$_{Ic}^{G}$ (MPa$\cdot\sqrt{m}$) & $\sigma_{max}^{\infty}$ (GPa) & K$_{Ic}^{\infty}$ (MPa$\cdot\sqrt{m}$) & K$_{Ic}^{G}$ (MPa$\cdot\sqrt{m}$) & $\sigma_{max}^{\infty}$ (GPa) & K$_{Ic}^{\infty}$ (MPa$\cdot\sqrt{m}$) & K$_{Ic}^{G}$ (MPa$\cdot\sqrt{m}$) & $\sigma_{max}^{\infty}$ (GPa)  \\
    \hline
    {$(0001)$$[10\overline{1}0]$ ($Nr.1$)} & {1.81 $\pm$ 0.03}  & {2.02} & {2.00 $\pm$ 0.77} & {2.03 $\pm$ 0.04} & {1.86} & {1.83 $\pm$ 0.65} & {2.75 $\pm$ 0.02} & {1.93} & {2.02 $\pm$ 0.64}\\
    {$(0001)$$[\overline{1}2\overline{1}0]$ ($Nr.2$)} & {2.35 $\pm$ 0.03}  & { 2.20} & {1.97 $\pm$ 0.49} & {2.08 $\pm$ 0.02} & {1.98} & {1.89 $\pm$ 0.22} & {2.89 $\pm$ 0.02} & {2.05} & {2.04 $\pm$ 0.31}\\
    {$(0001)$} & \multicolumn{3}{c}{$E_{surf}^{unrel}$: 4.02 (J/m$^{2}$)} & \multicolumn{3}{c}{$E_{surf}^{unrel}$: 3.68 (J/m$^{2}$)} & \multicolumn{3}{c}{$E_{surf}^{unrel}$: 3.61 (J/m$^{2}$)}\\
    \hline
    {$(\overline{1}2\overline{1}0)$$[10\overline{1}0]$ ($Nr.3$)} & {2.24 $\pm$ 0.02}  & {2.37} & {2.24 $\pm$ 0.63} & {1.74 $\pm$ 0.02 } & {2.18} & {2.21 $\pm$ 0.68} & {2.01 $\pm$ 0.02} & {2.20} & {2.27 $\pm$ 0.61} \\
    {$(\overline{1}2\overline{1}0)$$[0001]$ ($Nr.4$)} & {1.88 $\pm$ 0.02}  & { 2.29} & {2.22 $\pm$ 0.67} & {1.84 $\pm$ 0.02} & {2.15} & {1.75 $\pm$ 0.29} & {2.25 $\pm$ 0.05} & {2.18} & {1.61 $\pm$ 0.54}\\
    {$(\overline{1}2\overline{1}0)$} & \multicolumn{3}{c}{$E_{surf}^{unrel}$: 4.36 (J/m$^{2}$)} & \multicolumn{3}{c}{$E_{surf}^{unrel}$: 4.33 (J/m$^{2}$)} & \multicolumn{3}{c}{$E_{surf}^{unrel}$: 4.07 (J/m$^{2}$)}\\    
    \hline
    {$(10\overline{1}0)$$[0001]$ ($Nr.5$)} & {2.25 $\pm$ 0.02}  & {2.24} & {2.26 $\pm$ 0.69} & {1.74 $\pm$ 0.02} & {2.04} & {2.21 $\pm$ 0.68} & {2.74 $\pm$ 0.03} & {2.22} & {2.36 $\pm$ 0.47} \\
    {$(10\overline{1}0)$$[\overline{1}2\overline{1}0]$ ($Nr.6$)} & {2.24 $\pm$ 0.02}  & {2.16} & {1.97 $\pm$ 0.60} & {1.93 $\pm$ 0.02} & {2.01} & {2.11 $\pm$ 0.33} & {2.25 $\pm$ 0.02} & {2.20} & {2.26 $\pm$ 0.34}\\
    {$(10\overline{1}0)$} & \multicolumn{3}{c}{$E_{surf}^{unrel}$: 3.88 (J/m$^{2}$)} & \multicolumn{3}{c}{$E_{surf}^{unrel}$: 3.81 (J/m$^{2}$)} & \multicolumn{3}{c}{$E_{surf}^{unrel}$: 4.15 (J/m$^{2}$)}\\
\hline
\hline
\end{tabular}
}
\label{Comp}
\end{table*}

\begin{figure}[!h]
\centering
\includegraphics[width=0.5\textwidth]{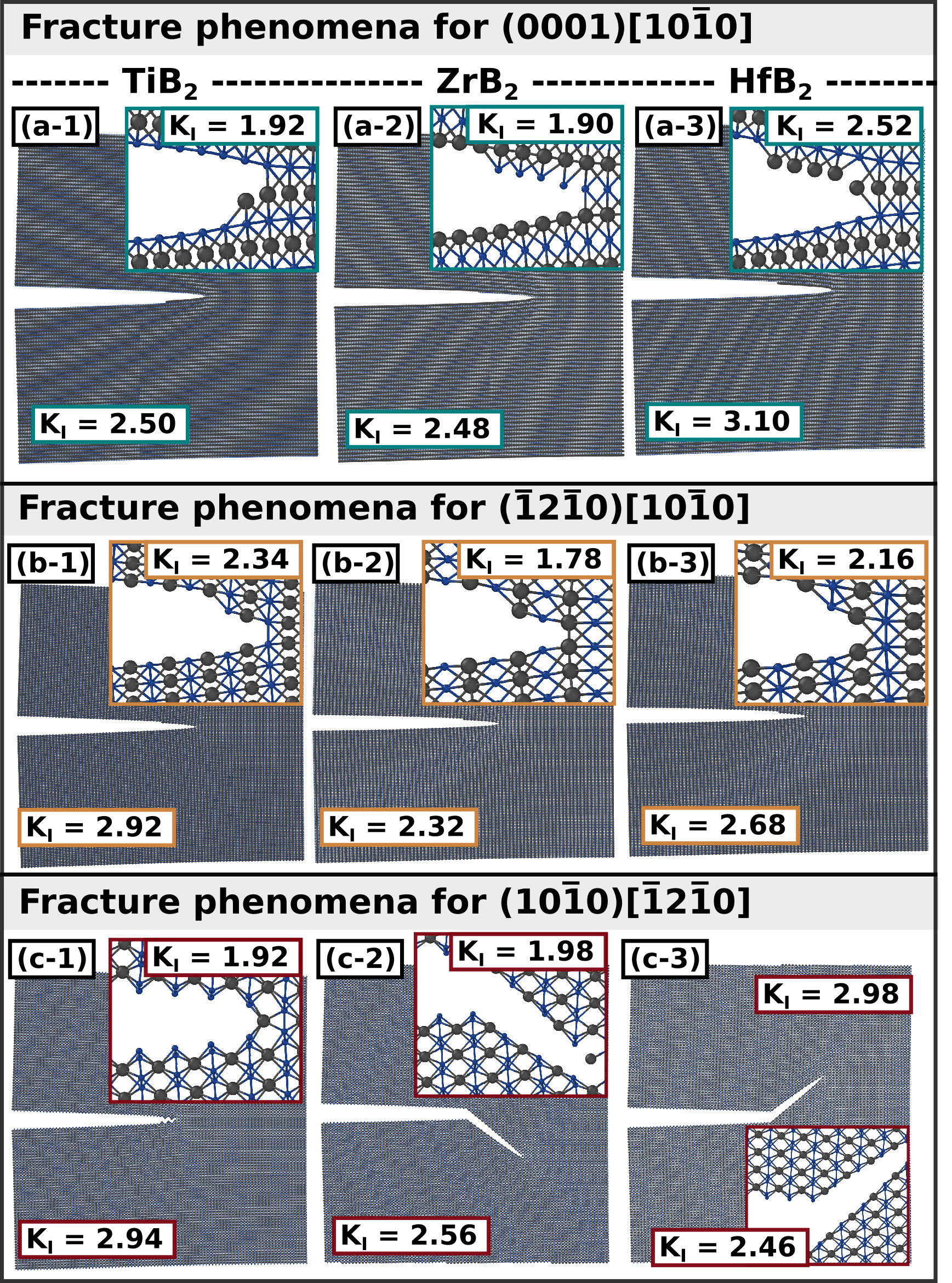}
\caption{\footnotesize 
{\bf{Fracture mechanisms in Group-IV TMB$_2$ compounds following crack initiation ($K_I \gtrsim K_{Ic}$)}}. Simulation snapshots are shown for one representative example per crack geometry: (a) $(0001)[10\overline{1}0]$ (Nr.~1), (b) $(\overline{1}2\overline{1}0)[10\overline{1}0]$ (Nr.~3), and (c) $(10\overline{1}0)[\overline{1}2\overline{1}0]$ (Nr.~6). Each panel compares TiB$_2$ (*-1), ZrB$_2$ (*-2), and HfB$_2$ (*-3) under pure Mode-I loading (the stress intensities $K_I$ are expressed in MPa$\cdot\sqrt{m}$). Each subpanel illustrates crack extension due to stress intensities well above $K_{Ic}$ ($K_I \gtrsim K_{Ic} + 0.5~\mathrm{MPa}\sqrt{\mathrm{m}}$), whereas the insets show magnifications of atomic configurations immediately after reaching $K_{Ic}$. Cracks are seen to either propagate along the initial plane or deflect depending on material and orientation. TM atoms (Ti, Zr, Hf) are depicted in dark gray; B atoms in light blue. The selected plate models have an area of $L^2 = 30$~nm $\times$ 30~nm = 900~nm$^2$. 
}
\label{mode1para}
\end{figure}

Fig.~\ref{mode1para} compares post-initiation fracture behavior in TiB\(_2\), ZrB\(_2\), and HfB\(_2\) for three representative geometries under Mode-I loading: (0001)[\(10\bar{1}0\)] (Nr.1), (\(1\bar{2}10\))[\(10\bar{1}0\)] (Nr.3), and (\(10\bar{1}0\))[\(11\bar{2}0\)] (Nr.6). Each subpanel displays the atomic configuration at the onset of crack propagation (\(K_{Ic}\)) and the resulting fracture morphology (crack wake) at \(K_I \gg K_{Ic}\). In all cases, fracture initiates through rupture of TM--B or B--B bonds (Fig.~\ref{Bond}), but the subsequent crack paths vary by material and geometry.

In diborides with native cracks on the (0001) and \((1\bar{2}10)\) surfaces, Mode-I loading results in crack propagation that remains confined to the original cleavage plane (see Fig.~\ref{mode1para}a and Fig.~\ref{mode1para}b). The surface energy differences reported in Tab.~\ref{SE} offer a plausible explanation for this behavior. Across all diboride systems, the ML-MS surface energies of the (0001) and \((1\bar{2}10)\) planes are similar to each other and consistently lower than that of the \((10\bar{1}0)\) surface. As a result, there is no strong energetic driving force to redirect the crack away from its initial orientation in these configurations.

More surprising material-specific behaviors emerge in Fig.~\ref{mode1para}c, where native cracks are positioned on the \((10\bar{1}0)\) planes -- identified as the highest-energy cleavage planes among those considered (Tab.~\ref{SE}). In ZrB\(_2\) and HfB\(_2\) (Fig.~\ref{mode1para}c-2 and c-3), these cracks readily deflect onto inclined \((1\bar{1}01)\) planes during Mode-I loading. The phenomenon is reminiscent of what is observed in TiN, where cracks on high-energy (111) surfaces extend onto lower-energy (001) planes (see Fig.~8a in Ref.~\cite{sangiovanni2024controlled}). Although surface energies for first-order pyramidal \((1\bar{1}01)\) planes are not available, the observed deflections suggest that these planes are energetically more favorable than \((10\bar{1}0)\) in ZrB\(_2\) and HfB\(_2\). Additional support comes from tensile simulations on small defect-free cells (Fig.~\ref{Vali}c), which show oblique fracture paths -- indicating an intrinsic tendency for crack redirection onto inclined planes that are presumably lower in energy. By contrast, in TiB\(_2\) (Fig.~\ref{mode1para}c-1), the crack propagates stably along the \((10\bar{1}0)\) plane without deviation. This suggests that any potential energetic advantage associated with switching to a pyramidal plane is not sufficient to activate deflection -- possibly due to competing crystallographic or mechanical constraints that favor continuation along the original cleavage plane.

In summary, although energetic arguments provide valuable guidance for interpreting fracture patterns, the examples shown in Fig.~\ref{mode1para}c highlight the inherent complexity of fracture phenomena. These cases illustrate how cleavage energetics and crystallographic constraints interact in a material-specific manner and underscore the need for direct atomistic simulations to fully resolve the evolution of crack paths.

To assess the relevance of our simulation results, we now compare them with experimental measurements of fracture toughness for TiB\(_2\), ZrB\(_2\), and HfB\(_2\). Reported values range between 1.8--6.8~MPa$\cdot\sqrt{m}$ for TiB\(_2\)~\cite{ferber1983effect,bhaumik2000synthesis,wang2002influence}, 2.2--5.0~MPa$\cdot\sqrt{m}$ for ZrB\(_2\)~\cite{monteverde2003advances,csanadi2024effect,swab2023mechanical}, and 2.8--7.2~MPa$\cdot\sqrt{m}$ for HfB\(_2\)~\cite{li2024synthesis,wang2015fabrication}. The calculated toughness values fall within these experimental ranges (see Fig.~\ref{mode1comp} and Tab.~\ref{Comp}), but direct comparison is complicated by the broad scatter in reported data. This variability arises from differences in sample preparation, microstructural characteristics (e.g., porosity, grain boundary density and grain sizes, and residual stresses), and the specific mechanical testing methods used. Furthermore, many experimental configurations involve mixed-mode loading -- typically a combination of Mode-I and Mode-II -- rather than ideal Mode-I conditions~\cite{maccagno1985brittle}. The role of mixed-mode effects is examined in the following section.

%% file: 3-Mix.tex
\section{Mixed Mode-I and Mode-II loading: example of TiB$_{2}$}

\begin{figure*}[!h]
\centering
\includegraphics[width=1.0\textwidth]{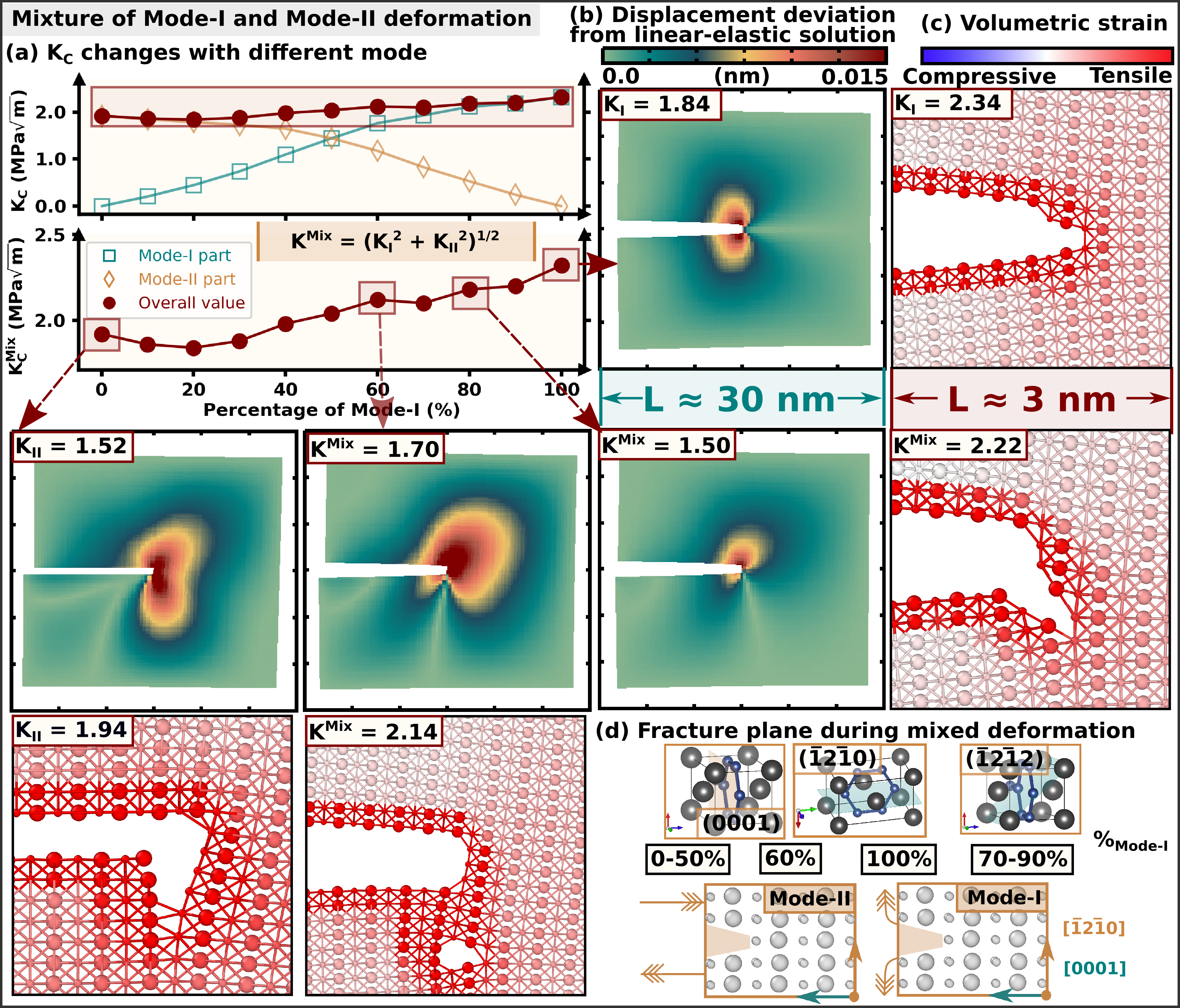}
\caption{\footnotesize 
{\bf{Fracture mechanisms under mixed Mode-I/Mode-II loading, as exemplified by TiB\(_2\) in the \((\overline{1}2\overline{1}0)[10\overline{1}0]\) (Nr.~3) crack geometry.}}
(a) Critical stress intensity values \(K_c\) (red) as a function of Mode-I contribution, decomposed into \(K_I\) (teal) and \(K_{II}\) (orange) components.  
(b) Maps of atomic displacement deviations from the corresponding linear-elastic solutions at a stress intensity \(K^{\text{mix}} \approx K_c - 0.02~\mathrm{MPa}\sqrt{\mathrm{m}}\), visualizing the effect of mixed-mode loading on lattice distortion.  
(c) Volumetric strain patterns near the crack tip under pure Mode-I and mixed-mode conditions, showing fracture behavior at \(K^{\text{mix}} \approx K_c + 0.02~\mathrm{MPa}\sqrt{\mathrm{m}}\). (d) Summary of fracture mechanisms as a function of Mode-I content: from basal cleavage under pure Mode-II, to (0001) and \((\overline{1}2\overline{1}2)\) plane fracture under mixed loading, to fully prismatic cleavage under pure Mode-I. Crystallographic directions \([0001]\), \([10\overline{1}0]\), and \([\overline{1}2\overline{1}0]\) are color-coded for reference. The stress intensity factors in (b) and (c), are expressed in \text{MPa}\(\cdot\sqrt{\text{m}}\).
}
\label{Mix-1}
\end{figure*}

\begin{figure*}[!h]
\centering
\includegraphics[width=1.0\textwidth]{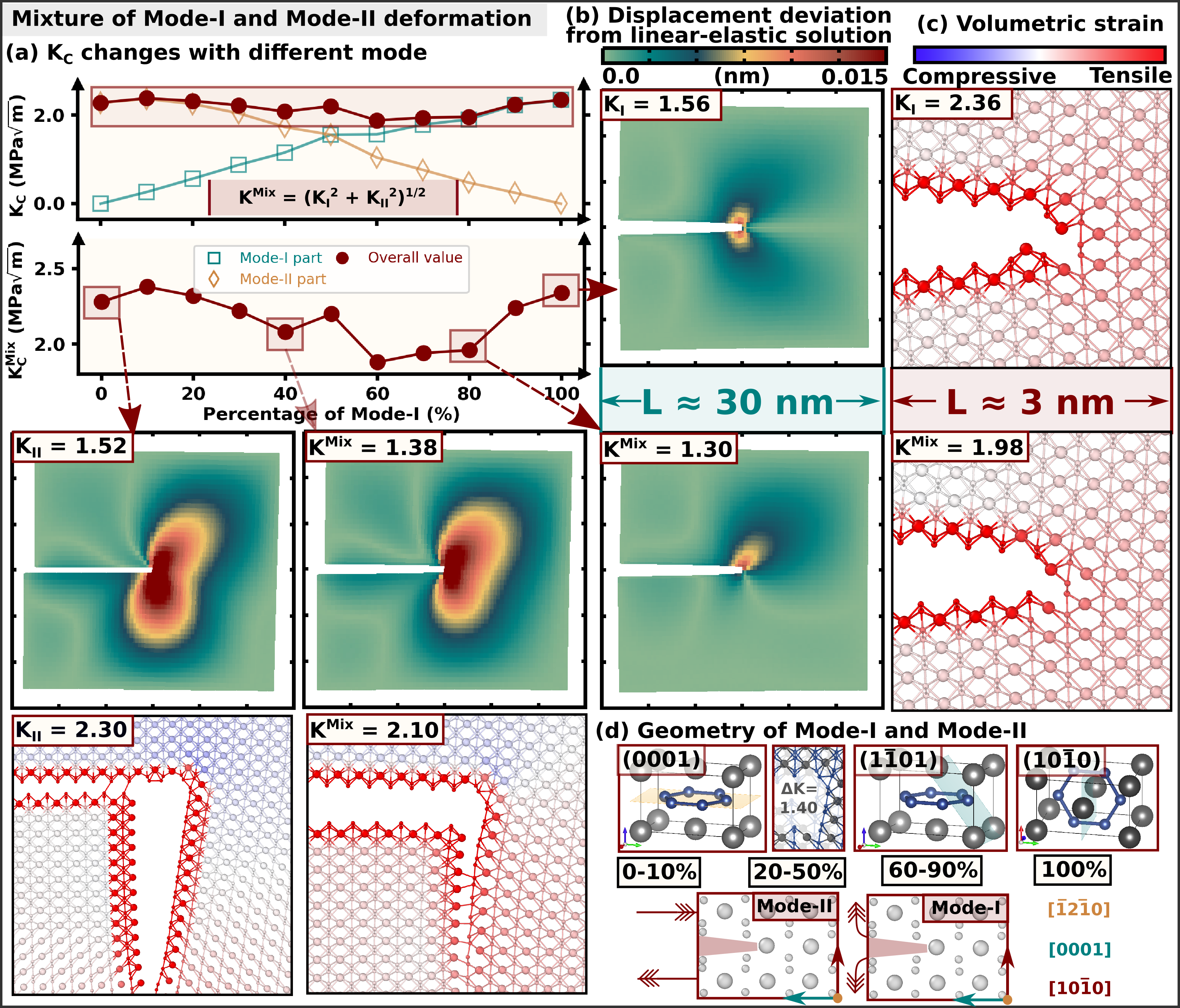}
\caption{\footnotesize 
{\bf{Fracture mechanisms under mixed Mode-I/Mode-II loading, as exemplified by TiB\(_2\) in the $(10\overline{1}0)$$[\overline{1}2\overline{1}0]$ ($Nr.6$) crack geometry.}} (a) Critical stress intensity values \(K_c\) (red) as a function of Mode-I contribution, decomposed into \(K_I\) (teal) and \(K_{II}\) (orange) components.  
(b) Maps of atomic displacement deviations from the corresponding linear-elastic solutions at a stress intensity \(K^{\text{mix}} \approx K_c - 0.02~\mathrm{MPa}\sqrt{\mathrm{m}}\), visualizing the effect of mixed-mode loading on lattice distortion.  
(c) Volumetric strain patterns near the crack tip under pure Mode-I and mixed-mode conditions, showing fracture behavior at \(K^{\text{mix}} \approx K_c + 0.02~\mathrm{MPa}\sqrt{\mathrm{m}}\). (d) Summary of fracture paths as a function of Mode-I content demonstrating four different mechanisms and three distinct fracture planes with varying ratios. Note that fracture at 20-50\% Mode-I loading initiates with lattice slip along the basal plane. Crystallographic directions \([0001]\), \([10\overline{1}0]\), and \([\overline{1}2\overline{1}0]\) are color-coded for reference. The stress intensity factors in (b) and (c), are expressed in \text{MPa}\(\cdot\sqrt{\text{m}}\).
}
\label{Mix-2}
\end{figure*}

Achieving a purely Mode-I fracture in mechanical testing experiments is uncommon due to material heterogeneity, sample geometry, or the specific loading configuration~\cite{royer1988study,de2009pure,jalayer2023novel}. Even minor misalignments, combined with microstructural features such as grain boundaries or phase interfaces, can introduce shear components that promote mixed-mode fracture~\cite{rizov2013mixed,braham2010laboratory,margevicius1999fracture}. In this section, we examine mixed Mode-I and Mode-II crack opening, using TiB\(_2\) as a representative system. Two crack geometries are considered for the main discussion: \((\overline{1}2\overline{1}0)[10\overline{1}0]\) (Nr.~3) and \((10\overline{1}0)[\overline{1}2\overline{1}0]\) (Nr.~6). Additionally, results for mixed loading obtained for the \((10\overline{1}0)[0001]\) (Nr.~5) and \((1\bar{2}10)[0001]\) (Nr.~4) cracked plate models are presented in the Supplemental Material, but briefly described below.

Fracture under mixed Mode-I and Mode-II loading can be described using energy-based criteria that account for the full crack-tip stress field. A well-known example is the strain energy density criterion by Sih and MacDonald~\cite{sih1974}, which evaluates the angular distribution of energy around the crack tip to predict both initiation and propagation direction. While rigorous, this approach requires evaluating the spatial dependence of stresses near the tip and is therefore complex to apply. Here, we adopt a common approximation in linear elastic fracture mechanics, taking the total driving force for crack growth as \( K^{\text{mix}} = \sqrt{K_I^2 + K_{II}^2} \) (see, e.g., Eq.~15 in Ref.~\cite{zou2025socket} and Eq.~7a in Ref.~\cite{predan2013stress}). That is, \(K^{\text{mix}}\) is the norm of a vector with orthogonal components \(K_I\) and \(K_{II}\). This definition provides a practical estimate of the overall stress intensity and enables consistent comparisons across different mixed-mode loading cases.

To analyze the influence of mixed Mode-I/Mode-II loading on fracture mechanisms in TiB\(_2\), we begin with simulations using the \((\overline{1}2\overline{1}0)[10\overline{1}0]\) cracked-plate model (Nr.~3). In this configuration, the relative contribution of Mode-I is varied in increments of 10\% from pure Mode-II (0\%) to pure Mode-I (100\%) (Fig.~\ref{Mix-1}a). At each step, the corresponding components \(K_I\) and \(K_{II}\) are computed according to Eqs.~\eqref{eq:Kmix}--\eqref{eq:KII} (see {\bf{Methods}} section), while \(K^{\text{mix}}\) is incremented in steps of 0.02~\text{MPa}\(\cdot\sqrt{\text{m}}\).


Fig.~\ref{Mix-1}a shows that the critical stress-intensity value for fracture initiation \(K_c^{\text{mix}}\) (hereafter abbreviated as \(K_c\)) generally increases with higher Mode-I contributions, rising from \(K_{IIc} = 1.92\)~\text{MPa}\(\cdot\sqrt{\text{m}}\) for pure Mode-II to \(K_{Ic} = 2.23\)~\text{MPa}\(\cdot\sqrt{\text{m}}\) for pure Mode-I. Interestingly, a slight drop in \(K_c\) is observed near the 20\% Mode-I / 80\% Mode-II ratio, where the toughness reaches a local minimum of 1.84~\text{MPa}\(\cdot\sqrt{\text{m}}\). This suggests that the \((\overline{1}2\overline{1}0)[10\overline{1}0]\) crack geometry in TiB\(_2\) is most susceptible to fracture under predominantly shear-dominated loading conditions with a modest tensile component. Figure~\ref{Mix-1}b displays the deviation in atomic displacements from the linear-elastic solution (see Ref.~\cite{andric2018atomistic} for technical details) at load levels near \(K_c\), capturing the evolution of the stress field. Distinct fracture responses emerge depending on the balance between shear and tensile loading. 

As shown in Figs.~\ref{Mix-1}c and \ref{Mix-1}d, crack propagation transitions through four different mechanisms depending on the mode ratio. Under pure Mode-I loading, the crack propagates predominantly along the original \((\overline{1}2\overline{1}0)\) fracture plane. As the Mode-I fraction decreases to intermediate levels, deflection occurs onto either the \((\overline{1}2\overline{1}2)\) or (0001) planes. In particular, at 70\% Mode-I content, the crack path shifts entirely to the \((\overline{1}2\overline{1}2)\) plane, marking a transition that begins with the activation of pyramidal slip and subsequently evolves into full oblique fracture. At 60\% Mode-I, the crack initially extends along the (0001) plane before redirecting to \((\overline{1}2\overline{1}0)\). For Mode-I fractions below 50\% -- including the pure Mode-II case -- the structure consistently fractures along the basal (0001) plane. This behavior closely resembles the crack growth mechanism observed in the \((1\bar{2}10)[0001]\) (Nr.~4) loading geometry. Further details are provided in the Supplementary Material.

The \((10\overline{1}0)[\overline{1}2\overline{1}0]\) (Nr.~6) deformation behavior (Fig.~\ref{Mix-2}a) exhibits a more intricate trend in \(K_c^{\text{mix}}\), including a minimum of 1.88~\text{MPa}\(\cdot\sqrt{\text{m}}\) at 60\% Mode-I. The toughness values for pure Mode-I and Mode-II are relatively similar--\(K_{Ic} = 1.72\)~\text{MPa}\(\cdot\sqrt{\text{m}}\) and \(K_{IIc} = 1.94\)~\text{MPa}\(\cdot\sqrt{\text{m}}\), respectively. Figure~\ref{Mix-2}b, showing the displacement deviation from the linear-elastic solution, highlights the evolving crack-tip stress fields. Combined with Figs.~\ref{Mix-2}c and \ref{Mix-2}d, these results reveal distinct fracture mechanisms as the loading mode varies.

Under pure Mode-I loading, the native \((10\overline{1}0)\) crack extends in a zigzag fashion along mirrored diagonal facets, maintaining an overall horizontal trajectory. Between 90\% and 60\% Mode-I, crack growth transitions to the \((1\overline{1}01)\) plane--consistent with the first-order pyramidal slip system and the preferred cleavage path in ZrB\(_2\) and HfB\(_2\) (see Fig.~\ref{mode1para}c). As the Mode-I fraction decreases further, between 50\% and 10\%, fracture begins with basal-plane slip followed by crack opening along the (0001) surface. A similar basal-slip mechanism has been observed in simple-shear simulations of single-crystal TMB\(_2\) (TM = Ti, Ta, W, Re) lattice models ~\cite{lin2024shear}. From 10\% Mode-I down to pure Mode-II, the crack opens directly along the (0001) surface, without preceding slip.

The evolution observed in the \((10\overline{1}0)[0001]\) (Nr.~5) configuration follows a more straightforward pattern. From pure Mode-I down to 90\% Mode-I, fracture proceeds along the \((10\overline{1}0)\) plane, though in a straight rather than zigzag path. At 80--70\% Mode-I, the crack shifts to the \((11\overline{2}0)\) plane, and for all lower Mode-I ratios -- including pure Mode-II -- fracture occurs through the \((1\overline{1}00)\) plane. These transitions suggest the activation of prismatic slip after reaching \(K_c\). Additional details are available in the Supplementary Material (Figs.~S4 and S6).

The fracture toughness values for the \((\overline{1}2\overline{1}0)[10\overline{1}0]\) and \((10\overline{1}0)[\overline{1}2\overline{1}0]\) configurations show only minor differences between \(K_{Ic}\) and \(K_{IIc}\). As seen in Fig.~\ref{Mix-1} and Fig.~\ref{Mix-2}, all four toughness values--corresponding to pure Mode-I and Mode-II loading for both geometries--cluster around 2~\text{MPa}\(\cdot\sqrt{\text{m}}\), indicating a consistent resistance to fracture regardless of loading mode or crack orientation. While this suggests similar macroscopic toughness under tension and shear, the underlying mechanisms differ markedly and confirm the strongly brittle nature of TiB\(_2\).

Under pure Mode-II loading, one would expect slip to initiate along the original crack plane--as seen in other hard ceramics. For instance, in TiN, atomistic simulations reveal that a \((110)[010]\) crack under pure Mode-II activates slip along the \(\{110\}\langle1\bar{1}0\rangle\) system, which lies within the original crack surface~\cite{sangiovanni2023descriptor}. In contrast, TiB\(_2\) shows no such response. Rather than slipping, TiB\(_2\) cracks along the orthogonal (0001) basal plane. This highlights TiB\(_2\)'s low propensity for shear-induced plasticity--even Mode-II loading is accommodated through brittle cleavage.

%% file: 3,5-Discussion.tex
\begin{figure}[!h]
\centering
\includegraphics[width=0.477\textwidth]{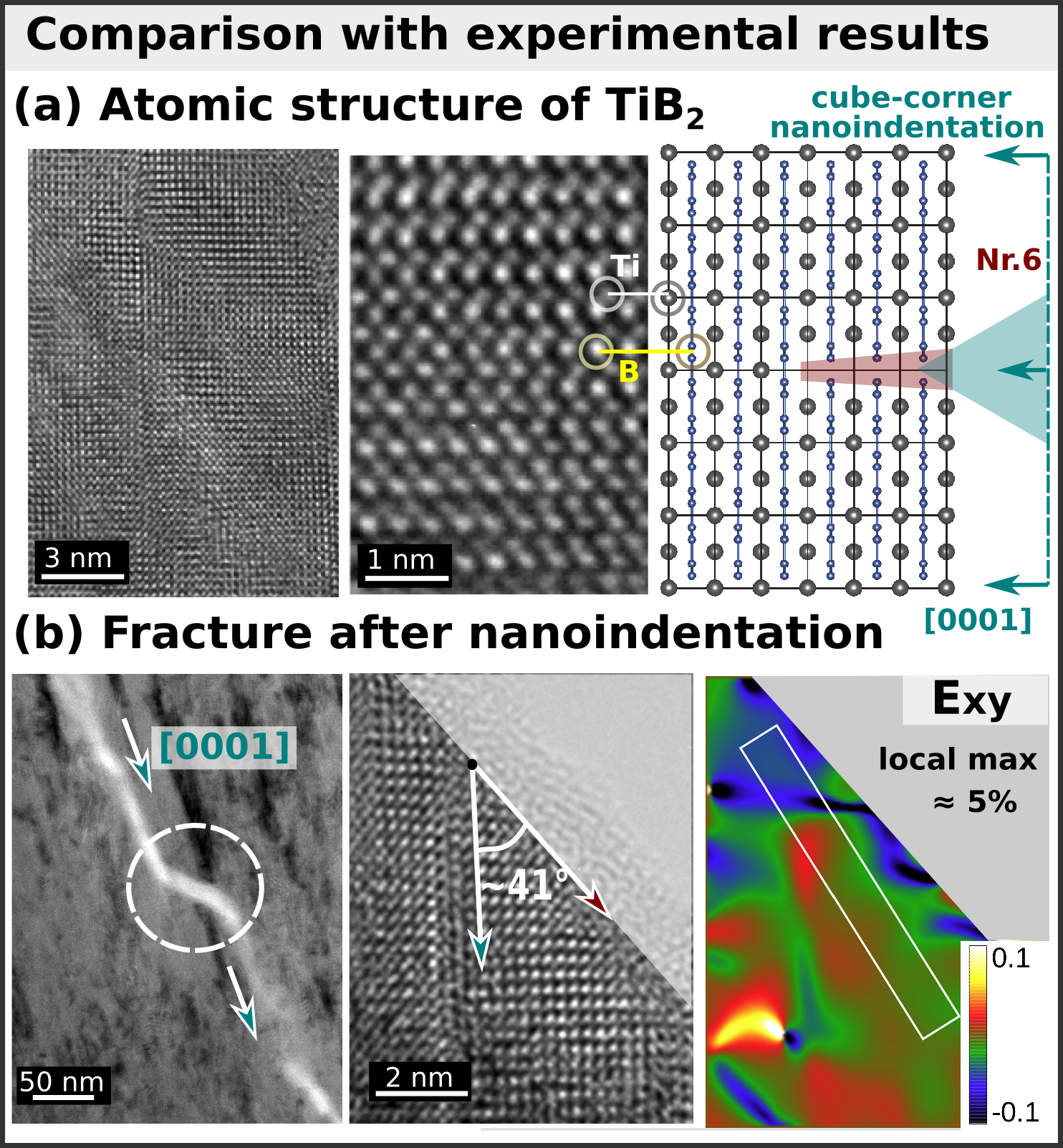}
\caption{\footnotesize 
{\bf{Comparison of simulation and experimental results for TiB$_{2}$.}}
(a) Atomic structure of TiB$_{2}$ thin film from HRTEM prior to nanoindentation, compared with the atomistic model used for pre-crack simulations, with the [0001] direction indicated.
The brighter pattern represents two nearest-neighbor B atoms merging together, while the relatively darker pattern corresponds to Ti atoms.
The contrast reversal of brightness is caused by the thicker TEM sample.
(b) HRTEM pattern after cube-corner nanoindentation, showing angle changes during fracture. 
The average shear strain, $E_{xy}$, is highlighted using the geometric phase analysis (GPA) pattern.
}
\label{Exp}
\end{figure}


To support the results of our atomistic simulations, we perform cube-corner nanoindentation experiments on TiB\(_2\) thin films with near 1:2 Ti-to-B stoichiometry. The film's microstructure before and after indentation is characterized using high-resolution transmission electron microscopy (HRTEM). The as-deposited samples exhibit [0001] orientation (Fig.~\ref{Exp}a), consistent with the typical growth direction of TMB\(_2\) thin films.

Cube-corner indentation is known to generate a combination of Mode-I (opening) and Mode-II (shearing) stresses~\cite{gupta2020fracture}. Post-mortem HRTEM analysis of the [0001]-oriented TiB\(_2\) films reveals that cracks initiate from the indentation site along a prismatic plane and subsequently deflect onto an inclined plane at approximately \(40^\circ\) (Fig.~\ref{Exp}b). This oblique fracture trajectory suggests that crack propagation occurs under mixed-mode loading conditions. To interpret this behavior, we compare the experimental observations with our ML-MS simulations of defective lattices containing native prismatic cracks -- that is, orthogonal to the basal plane -- shown in Figs.~\ref{Mix-1} and \ref{Mix-2}.

As illustrated in Fig.~\ref{mode1para}b-1 and Fig.~\ref{mode1para}c-1, cracks in TiB\(_2\) loaded under pure Mode-I conditions propagate straight along the original prismatic plane--a response inconsistent with crack deflection observed in the experimental test. In contrast, simulations under mixed Mode-I/Mode-II loading (Figs.~\ref{Mix-1} and \ref{Mix-2}) reveal crack redirection toward inclined pyramidal planes, especially when the Mode-I contribution lies between 60\% and 90\% (see Figs.~\ref{Mix-1}d and \ref{Mix-2}d). In particular, simulations of \((10\overline{1}0)[\overline{1}2\overline{1}0]\) cracked-plate models under \(<40\%\) Mode-II loading component show diagonal crack propagation along the \((1\overline{1}01)\) plane--closely matching the experimental deflection angle and aligning with the relatively low energetic cost of pyramidal slip in AlB\(_2\)-type materials~\cite{hunter2016investigations,fuger2022anisotropic}.

Additionally, geometric phase analysis (GPA) of the post-mortem microstructure reveals a local maximum shear strain of approximately 5\% near the crack tip, consistent with a mixed-mode fracture mechanism involving Mode-II contribution. Taken together, these findings suggest that the experimentally observed crack deflection originates from mixed-mode loading during nanoindentation and reinforce the relevance of our atomistic simulations for predicting and interpreting fracture patterns in brittle ceramics.

The fracture toughness measured for the same TiB\(_2\) sample by microcantilever bending yielded a value of 3.08~MPa\(\cdot\sqrt{\text{m}}\)~\cite{fuger2023tissue}, approximately 40\% higher than both the extrapolated \(K_{Ic}^\infty\) and the Griffith-based estimate \(K_{Ic}^G\). This difference likely arises from several factors. First, the experimental measurement reflects the response of a columnar polycrystalline matrix, where microstructural features such as grain boundaries and local residual stresses (up to $\approx$ 3~GPa compressive residual stress in this series of TiB$_{2}$ sample~\cite{hirle2025mechanical}) can enhance resistance to crack propagation -- effects that are not captured in our idealized atomistic models. Second, the geometry of a notched microcantilever -- with notch widths on the order of tens of nanometers~\cite{fuger2023tissue} -- differs significantly from the atomically sharp cracks modeled in simulations. Prior atomistic work on TiN(001) -- a similarly brittle ceramic -- showed that fracture toughness increases with notch width and saturates at a value roughly 20\% above the atomically sharp case, once the notch spans just a few atomic layers~\cite{sangiovanni2023descriptor}. This suggests that toughness values obtained from finite-width experimental notches are inherently higher than those predicted for atomically sharp cracks.

A further contribution to the discrepancy may stem from how fracture initiation is defined in our simulations. To ensure consistency across all crack geometries and materials, we identify fracture onset with the rupture of the first chemical bond at the crack tip. While this criterion is unambiguous and reproducible, it neglects the role of lattice trapping and other atomistic effects that can locally stabilize the crack front, effectively delaying propagation. As a result, the extrapolated \(K_{Ic}^\infty\) values may modestly underestimate the effective macroscopic toughness, particularly for configurations prone to such trapping phenomena.

Looking ahead, a key challenge for MLIP-based atomistic modeling is to incorporate simplified but representative features of real microstructures. These may include common point defects such as boron or metal vacancies~\cite{hu2024influence,glechner2022influence}, extended planar defects like anti-phase boundaries~\cite{palisaitis2022nature,palisaitis2021unpaired}, or amorphous-like B-rich columns~\cite{thornberg2020microstructure}. However, it is not sufficient to introduce defects arbitrarily. To enable meaningful comparison with experimental data, one must first identify which defect types and distributions are relevant for the adopted synthesis conditions, and then construct ad hoc atomistic models that reflect those microstructural features. As emphasized in Ref.~\cite{andric2018atomistic}, these models must be built with care to avoid unphysical artifacts. For example, placing a native dislocation too close to the crack plane can significantly alter the strain field near the crack tip, biasing the computed toughness and limiting the generality of the results. Ultimately, by combining process-aware defect modeling with rigorous \(K\)-controlled simulations, it may become possible to quantitatively link intrinsic fracture properties to those measured in structurally complex, real-world TiB\(_2\)-based ceramics.

%% file: 4-Conclusion.tex
To investigate the intrinsic fracture properties of Group-IV transition metal diborides (TMB\(_2\), TM = Ti, Zr, Hf), we performed \(K\)-controlled molecular statics simulations on pre-cracked lattice models using machine-learning interatomic potentials (MLIPs). The potentials were validated following a protocol analogous to our previous work~\cite{lin2024machine}, ensuring that surface energies, elastic constants, and theoretical tensile strengths reproduce DFT and \textit{ab initio} MD results.

Our analysis focused on six low-index atomically sharp crack geometries under pure Mode-I loading: (0001)$[10\overline{1}0]$, (0001)$[\overline{1}2\overline{1}0]$, $(\overline{1}2\overline{1}0)$$[10\overline{1}0]$, $(\overline{1}2\overline{1}0)$[0001], $(10\overline{1}0)$[0001], and $(10\overline{1}0)$$[\overline{1}2\overline{1}0]$, where (hklm) denotes the crack surface and [h'k'l'm'] the crack line or crack front direction. The macroscale fracture toughness \(K_{Ic}^\infty\) and fracture strength \(\sigma_\text{max}^\infty\) were obtained by extrapolating results from different plate sizes. While all diboride systems exhibit similar fracture strengths (\(\approx\)2.0~MPa$\cdot\sqrt{m}$), the toughness ranked as: HfB\(_2\) (\(\approx\)2.7~MPa$\cdot\sqrt{m}$) $>$ TiB\(_2\) (\(\approx\)2.3~MPa$\cdot\sqrt{m}$) $>$ ZrB\(_2\) (\(\approx\)1.8~MPa$\cdot\sqrt{m}$), with minor deviations in two geometries. The predicted values fall within the ranges reported experimentally: 1.8--6.8~MPa$\cdot\sqrt{m}$ for TiB\(_2\)~\cite{ferber1983effect,bhaumik2000synthesis,wang2002influence}, 2.2--5.0~MPa$\cdot\sqrt{m}$ for ZrB\(_2\)~\cite{monteverde2003advances,csanadi2024effect,swab2023mechanical}, and 2.8--7.2~MPa$\cdot\sqrt{m}$ for HfB\(_2\)~\cite{li2024synthesis,wang2015fabrication}. Deviations from experimental values can be attributed to the absence of microstructural features in simulations (e.g., grain boundaries, residual stress), wider notches used in experiments~\cite{fuger2023tissue}, and our strict definition of fracture onset -- based on the first bond rupture -- which overlooks effects like lattice trapping that can delay crack advance.

We further explored the influence of mixed Mode-I/Mode-II loading on fracture resistance \(K_c\) and crack path evolution in TiB\(_2\), focusing on the (\(\overline{1}2\overline{1}0\))$[10\overline{1}0]$ and (10\(\overline{1}0\))$[\overline{1}2\overline{1}0]$ geometries. Simulations showed that mixed-mode loading can significantly alter fracture trajectories and reduce \(K_c\). In particular, \(K_c\) minima emerged under specific shear/tensile combinations, with cracks deflecting toward inclined pyramidal planes. This behavior was confirmed experimentally by cube-corner nanoindentation on [0001]-oriented TiB\(_2\) thin films, which exhibited oblique crack propagation at \(\approx 40^\circ\). Geometric phase analysis revealed local shear strain of \(\approx 5\%\), consistent with mixed-mode loading and supporting our atomistic predictions.

Taken together, these results demonstrate that \(K\)-controlled ML-MS simulations offer a predictive, atomistically detailed framework for evaluating fracture properties in brittle ceramics. By resolving how loading mode, crystallographic orientation, and local structure influence crack evolution, our approach provides both quantitative and mechanistic insight. Although finite-temperature effects were not explicitly modeled, the agreement between ML-MD and AIMD in prior validation supports future extensions of this framework to dynamic and environment-sensitive fracture phenomena. This paves the way toward systematic prediction of fracture-related descriptors and mechanisms in complex ceramic systems.

%% file: 5-Methods.tex
\subsection*{{\textit{Ab initio}} calculations}
Finite-temperature Born-Oppenheimer {\it{ab initio}} molecular dynamics ({\it{ab initio}} MD) calculations were carried out using VASP~\cite{VASP-1} together with the projector augmented wave (PAW)~\cite{VASP-2} method and the Perdew-Burke-Ernzerhof exchange-correlation functional revised for solids (PBEsol)~\cite{PBEsol}.
The plane-wave cut-off energies of 300~eV and $\Gamma$-point sampling of the reciprocal space were employed.

{\bf{Structural models}} of TMB$_2$:s, TM$=$(Ti, Zr, Hf) were based on the $\alpha$ polymorph, adopting AlB$_2$-type (P6/mmm) phase~\cite{leiner2023energetics,hu2025microstructure}.
The hexagonal unit cells were orthogonalized using the following crystallographic orientations: $x\parallel[10\overline{1}0]$, $y\parallel[\overline{1}2\overline{1}0]$, $z\parallel[0001]$.
All {\it{ab initio}} MD calculations were conducted using 720-atom supercells (240~TM$+$480~B) with dimensions of $\approx(1.5 \times 1.6 \times 2.6)~\mathrm{nm}^3$.
The supercells were {{equilibrated}} at target temperature (300~K and 1200~K) through a two-step process: 
(i) a 10\;ps isobaric-isothermal (NpT) equilibration simulation using the Parrinello-Rahman barostat~\cite{NPT} and the Langevin thermostat; 
(ii) a 2--4\;ps simulation with the canonical (NVT) ensemble based on Nosé-Hoover thermostat, imposing the time-averaged lattice parameters obtained from the equilibration stage (i).

{\bf{Room-temperature elastic constants}}, $C_{ij}$, were determined following  Ref.~\cite{sangiovanni2021temperature}, i.e., obtained from a second-order polynomial fit of stress/strain data from the [0001], [10$\overline{1}0$], and [$\overline{1}2\overline{1}0$] tensile simulations (used to derive $C_{11}$, $C_{12}$, $C_{13}$, $C_{33}$), and the (0001)$[\overline{1}2\overline{1}0]$, $(10\overline{1}0)[\overline{1}2\overline{1}0]$ and $(10\overline{1}0)[0001]$ shear simulations (used to derive $C_{44}$).
Strains ranging from 0 to 4\% were considered.
Stress tensor components were calculated by averaging data over the final 0.5~ps of each simulation.
{\bf{Zero Kelvin elastic constants}} were also calculated using the stress-strain method, with the same energy convergence criteria as the {\it{ab initio}} MD simulations but smaller strain ($\leq$ 1\%).
The methodology is consistent with our previous work~\cite{lin2024machine} on $\alpha$-TiB$_2$.
The {\bf{{surface energies}}} were calculated at zero Kelvin using 60-atom TMB$_2$ supercells (${3\times3\times1}$ \textbf{k}-mesh and cut-off energy of 300~eV) together with a {12~\AA} vacuum layer.

\subsection*{Molecular statics/dynamics with MLIPs (ML-MS/MD)}
ML-MS calculations were conducted using the LAMMPS code~\cite{LAMMPS} interfaced with the {\texttt{mlip-2}} package~\cite{MLIP} enabling the usage of MTP-type MLIPs.
Active learning has been performed using the concept of extrapolation grade,  MV~\cite{MV}, which also served to assess the reliability during MD simulations with the trained MLIPs.
For MLIP {\bf{validation}} purposes, we performed MD simulations of uniaxial deformation and calculated elastic constants. 
Computational setup for equilibration and tensile tests at the atomic scale was designed to closely match the corresponding {\it{ab initio}} MD simulations. 
For zero-Kelvin elastic constant validation, to minimize variables and maintain accuracy, we conducted MD calculations at 10~K instead of using MS simulations.
A detailed discussion of MD validation is provided in the supplementary material.
The zero-Kelvin surface energies were validated using ML-MS with 12~{\AA} vacuum, the same as in DFT calculations.

{\bf{$K$-controlled MS simulations}} were utilized to evaluate the effective resistance to brittle cleavage, using cracked-plate models of sizes reaching $\approx{10^6}$ atoms ($A=L^{2}\approx250^{2}$nm$^{2}$, where $A$ is the plate area and $L$ is the lateral size).
Theory and methods followed Refs.~\cite{andric2017new,andric2018atomistic,sangiovanni2023descriptor,sangiovanni2024controlled}. 
We employed square TMB$_{2}$ plates (with area of $L^2$) with all the possible geometries for both Mode-I and Mode-II loadings, i.e., (0001)$[10\overline{1}0]$ ($Nr.1$), (0001)$[\overline{1}2\overline{1}0]$ ($Nr.2$), $(\overline{1}2\overline{1}0)$$[10\overline{1}0]$ ($Nr.3$), $(\overline{1}2\overline{1}0)$[0001] ($Nr.4$), $(10\overline{1}0)$[0001] ($Nr.5$), $(10\overline{1}0)$$[\overline{1}2\overline{1}0]$ ($Nr.6$).
The supercells were periodic along the crack-front direction with thickness of approximately 0.5~nm. 

Atoms in the frame region centered at the crack tip are incrementally displaced by applying increasing values of the stress intensity factors \(K_I\), \(K_{II}\), and \(K_{\text{mix}}\), using a step size of 0.02~\text{MPa}\(\cdot\sqrt{\text{m}}\). For mixed-mode loading, the components are calculated as follows:
\begin{subequations} \label{eq:Kdefs}
\begin{align}
 K_{\text{mix}} &= \sqrt{K_I^2 + K_{II}^2} \label{eq:Kmix}, \tag{1.a} \\
K_I &= K_{\text{mix}} \cdot \frac{a}{\sqrt{a^2 + b^2}} \label{eq:KI}, \tag{1.b} \\
K_{II} &= K_{\text{mix}} \cdot \frac{b}{\sqrt{a^2 + b^2}} \label{eq:KII}, \tag{1.c}
\end{align}
\end{subequations}
where \(a\) and \(b\) represent the prescribed Mode-I and Mode-II percentages, such that \(a + b = 100\).

All, except frame atoms, were relaxed with the conjugate-gradient algorithm at each $K$ increment, with tolerances set to 10$^{-14}$ for the relative change in energy and 10$^{-14}$ eV/{\AA} for forces. 
The simulations were carried out with atomically sharp cracks (deleting half monolayer), where interactions between atoms on opposite sides of the crack plane were screened over $\approx{1}$~nm. 
Additionally, the cracked plates are constructed with an equal number of atomic layers above and below the crack plane. 

The {\bf{Griffith fracture toughness}}, $K_{Ic}^{G}$, was derived from~\cite{ting1996anisotropic}:
\begin{equation}
K_{\mathrm{Ic}}^{\mathrm{G}}=\left[2 E_{\text {surf}}^{\text {unrel }} \Lambda_{22}^{-1}\right]^{1 / 2},
\label{Gri}
\end{equation}
where $E_{surf}^{unrel}$ represents the unrelaxed surface energy and $\overline{\bar{\Lambda}}$ is the Stroh energy tensor, calculated from the elastic tensor.

\subsection*{Experimental methods}

A TiB$_{2.04}$ coating with near 1:2 stoichiometry and approximately 2 $\mu$m thickness was provided for analysis. 
The coating was deposited using an in-house DC balanced magnetron sputtering system, employing a 6-inch, powder-metallurgically produced TiB$_{2}$/C (99/1 wt.\%) target (>99.6\% purity). 
The complete deposition procedure is detailed in Ref.~\cite{fuger2023tissue}. 
Nanoindentation tests were performed using an CSIRO UMIS indenter equipped with a cube-corner diamond tip to induce controlled cracking. 
A total of 17 indents were applied at peak loads (F$_{m}$) ranging from 50 to 450 mN. 
The tests followed a force-controlled loading-unloading cycle. 
A cross-sectional TEM lamella was extracted from a radial crack tip using a Thermo Scientific Scios 2 DualBeam FIB-SEM system. 
Following a conventional FIB milling and lift-out procedure as presented in Ref.~\cite{schaffer2012sample}, an 8 $\mu$m thick tungsten protection layer was deposited over the region of interest to prevent milling damage. 
Initially, a 2~$\mu$m thick cross-sectional lamella was prepared, which was then refined by sequential ion milling steps to approximately 100 nm. 
Final cleaning steps at 2 kV and 27 pA, followed by Ar ion milling at 0.5 kV using a Gatan PIPS II system, further reduced the thickness to <75 nm, allowing high-resolution TEM analysis.


The cross-sectional transmission electron microscopy (TEM) specimens were prepared using an FEI Quanta 200 3D DBFIB. 
A 200~kV field emission TEM (JEOL 2100F) equipped with an image-side spherical aberration ($C_S$)-corrector was used in the high-resolution TEM (HRTEM) study, demonstrating a resolution of 1.2~{\AA} at 200~kV.
The aberration coefficient was set close to zero, under which the HRTEM images were taken under slightly over-focus conditions (close to the Scherzer defocus). 
A CCD Orius camera is used to record HRTEM images, where image sizes are 2048~pixels $\times$~1336 pixels.  
The strain fields in TiB$_{2}$ were calculated based on the $C_S$-corrected HRTEM images by the geometric phase analysis (GPA) method. 
According to the GPA algorithm, the displacement fields can be obtained by selecting two non-collinear Bragg vectors in the power spectrum generated from a HRTEM image.

%% file: main.bbl
\begin{thebibliography}{10}

\bibitem{griffith1921vi}
A.~A. Griffith, ``Vi. the phenomena of rupture and flow in solids,'' {\em Proc R Soc Lond A Math Phys Sci}, vol.~221, no.~582-593, pp.~163--198, 1921.

\bibitem{irwin1957analysis}
G.~R. Irwin, ``Analysis of stresses and strains near the end of a crack traversing a plate,'' {\em Journal of Applied Mechanics}, vol.~24, pp.~361--364, 06 2021.

\bibitem{nose1988evaluation}
T.~Nose and T.~FUJII, ``Evaluation of fracture toughness for ceramic materials by a single-edge-precracked-beam method,'' {\em J. Am. Ceram. Soc.}, vol.~71, no.~5, pp.~328--333, 1988.

\bibitem{anderson1985elastic}
T.~L. Anderson, M.~Dawes, and H.~McHenry, {\em Elastic-plastic fracture toughness tests with single-edge notched bend specimens}.
\newblock ASTM International, 1985.

\bibitem{shaji2003plane}
E.~M. Shaji, S.~R. Kalidindi, R.~D. Doherty, and A.~S. Sedmak, ``Plane strain fracture toughness of {MP35N} in aged and unaged conditions measured using modified {CT} specimens,'' {\em Mater. Sci. Eng. A}, vol.~340, no.~1-2, pp.~163--169, 2003.

\bibitem{underwood1984review}
J.~Underwood, S.~Freiman, F.~Baratta, {\em et~al.}, ``A review of chevron-notched fracture specimens,'' in {\em Chevron-notched Specimens, Testing and Stress Analysis: A Symposium}, vol.~855, p.~5, ASTM International, 1984.

\bibitem{kolhe1998effects}
R.~Kolhe, C.-Y. Hui, and A.~T. Zehnder, ``Effects of finite notch width on the fracture of chevron--notched specimens,'' {\em Int. J. Fract.}, vol.~94, pp.~189--198, 1998.

\bibitem{gludovatz2010fracture}
B.~Gludovatz, S.~Wurster, A.~Hoffmann, and R.~Pippan, ``Fracture toughness of polycrystalline tungsten alloys,'' {\em Int. J. Refract. Met. Hard Mater.}, vol.~28, no.~6, pp.~674--678, 2010.

\bibitem{launey2009fracture}
M.~E. Launey and R.~O. Ritchie, ``On the fracture toughness of advanced materials,'' {\em Adv. Mater.}, vol.~21, no.~20, pp.~2103--2110, 2009.

\bibitem{zeng2010modeling}
X.~Zeng and A.~Hartmaier, ``Modeling size effects on fracture toughness by dislocation dynamics,'' {\em Acta Mater.}, vol.~58, no.~1, pp.~301--310, 2010.

\bibitem{moller2014fracture}
J.~J. M{\"o}ller and E.~Bitzek, ``Fracture toughness and bond trapping of grain boundary cracks,'' {\em Acta Mater.}, vol.~73, pp.~1--11, 2014.

\bibitem{shimokawa2011roles}
T.~Shimokawa, M.~Tanaka, K.~Kinoshita, and K.~Higashida, ``Roles of grain boundaries in improving fracture toughness of ultrafine-grained metals,'' {\em Phys. Rev. B Condens. Matter}, vol.~83, no.~21, p.~214113, 2011.

\bibitem{samborski2010dynamic}
S.~Samborski and T.~Sadowski, ``Dynamic fracture toughness of porous ceramics,'' {\em J. Am. Ceram. Soc.}, vol.~93, no.~11, pp.~3607--3609, 2010.

\bibitem{kleebe1999microstructure}
H.-J. Kleebe, G.~Pezzotti, and G.~Ziegler, ``Microstructure and fracture toughness of {Si}$_{3}${N}$_{4}$ ceramics: combined roles of grain morphology and secondary phase chemistry,'' {\em J. Am. Ceram. Soc.}, vol.~82, no.~7, pp.~1857--1867, 1999.

\bibitem{zhang2021correlating}
Z.~Zhang, A.~Ghasemi, N.~Koutn{\'a}, Z.~Xu, T.~Gr{\"u}nst{\"a}udl, K.~Song, D.~Holec, Y.~He, P.~H. Mayrhofer, and M.~Bartosik, ``Correlating point defects with mechanical properties in nanocrystalline {TiN} thin films,'' {\em Mater. Des.}, vol.~207, p.~109844, 2021.

\bibitem{ma2021situ}
L.~Ma, A.-L. Fauchille, M.~R. Chandler, P.~Dowey, K.~G. Taylor, J.~Mecklenburgh, and P.~D. Lee, ``In-situ synchrotron characterisation of fracture initiation and propagation in shales during indentation,'' {\em Energy}, vol.~215, p.~119161, 2021.

\bibitem{fadenberger2010situ}
K.~Fadenberger, I.~E. Gunduz, C.~Tsotsos, M.~Kokonou, S.~Gravani, S.~Brandstetter, A.~Bergamaschi, B.~Schmitt, P.~H. Mayrhofer, C.~C. Doumanidis, {\em et~al.}, ``In situ observation of rapid reactions in nanoscale {Ni}--{Al} multilayer foils using synchrotron radiation,'' {\em Appl. Phys. Lett.}, vol.~97, no.~14, 2010.

\bibitem{mishin2021machine}
Y.~Mishin, ``Machine-learning interatomic potentials for materials science,'' {\em Acta Mater.}, vol.~214, p.~116980, 2021.

\bibitem{dragoni2018achieving}
D.~Dragoni, T.~D. Daff, G.~Cs{\'a}nyi, and N.~Marzari, ``Achieving {DFT} accuracy with a machine-learning interatomic potential: Thermomechanics and defects in bcc ferromagnetic iron,'' {\em Phys. Rev. Mater.}, vol.~2, no.~1, p.~013808, 2018.

\bibitem{mueller2020machine}
T.~Mueller, A.~Hernandez, and C.~Wang, ``Machine learning for interatomic potential models,'' {\em J. Chem. Phys.}, vol.~152, no.~5, 2020.

\bibitem{zhang2024efficiency}
L.~Zhang, G.~Cs{\'a}nyi, E.~van~der Giessen, and F.~Maresca, ``Efficiency, accuracy, and transferability of machine learning potentials: Application to dislocations and cracks in iron,'' {\em Acta Mater.}, vol.~270, p.~119788, 2024.

\bibitem{behler2016perspective}
J.~Behler, ``Perspective: Machine learning potentials for atomistic simulations,'' {\em J. Chem. Phys.}, vol.~145, no.~17, 2016.

\bibitem{deringer2019machine}
V.~L. Deringer, M.~A. Caro, and G.~Cs{\'a}nyi, ``Machine learning interatomic potentials as emerging tools for materials science,'' {\em Adv. Mater.}, vol.~31, no.~46, p.~1902765, 2019.

\bibitem{zuo2020performance}
Y.~Zuo, C.~Chen, X.~Li, Z.~Deng, Y.~Chen, J.~Behler, G.~Cs{\'a}nyi, A.~V. Shapeev, A.~P. Thompson, M.~A. Wood, {\em et~al.}, ``Performance and {C}ost {A}ssessment of {M}achine {L}earning {I}nteratomic {P}otentials,'' {\em J. Phys. Chem. A}, vol.~124, no.~4, pp.~731--745, 2020.

\bibitem{smith2017ani}
J.~S. Smith, O.~Isayev, and A.~E. Roitberg, ``{ANI-1}: an extensible neural network potential with {DFT} accuracy at force field computational cost,'' {\em Chem. Sci.}, vol.~8, no.~4, pp.~3192--3203, 2017.

\bibitem{shapeev2020elinvar}
A.~V. Shapeev, E.~V. Podryabinkin, K.~Gubaev, F.~Tasn{\'a}di, and I.~A. Abrikosov, ``Elinvar effect in $\beta$-{Ti} simulated by on-the-fly trained moment tensor potential,'' {\em New J. Phys.}, vol.~22, no.~11, p.~113005, 2020.

\bibitem{MLIP}
I.~S. Novikov, K.~Gubaev, E.~V. Podryabinkin, and A.~V. Shapeev, ``The $\mathrm{MLIP}$ package: moment tensor potentials with $\mathrm{MPI}$ and active learning,'' {\em Mach. learn.: sci. technol.}, vol.~2, no.~2, p.~025002, 2020.

\bibitem{lin2024machine}
S.~Lin, L.~Casillas-Trujillo, F.~Tasn{\'a}di, L.~Hultman, P.~H. Mayrhofer, D.~G. Sangiovanni, and N.~Koutn{\'a}, ``Machine-learning potentials for nanoscale simulations of tensile deformation and fracture in ceramics,'' {\em Npj Comput. Mater.}, vol.~10, no.~1, p.~67, 2024.

\bibitem{lin2024shear}
S.~Lin, D.~Holec, D.~Sangiovanni, T.~Leiner, L.~Hultman, P.~Mayrhofer, and N.~Koutná, ``{Shear-activated phase transformations of diborides via machine-learning potential molecular dynamics},'' {\em Preprint}, vol.~10.21203/rs.3.rs-5327540/v1, 2024.

\bibitem{koutna2025machine}
N.~Koutn{\'a}, S.~Lin, L.~Hultman, D.~G. Sangiovanni, and P.~H. Mayrhofer, ``Machine-learning potentials for structurally and chemically complex {MAB} phases: strain hardening and ripplocation-mediated plasticity,'' {\em Preprint}, vol.~dx.doi.org/10.2139/ssrn.5187025, 2025.

\bibitem{sevik2022high}
C.~Sevik, J.~Bekaert, M.~Petrov, and M.~V. Milo{\v{s}}evi{\'c}, ``High-temperature multigap superconductivity in two-dimensional metal borides,'' {\em Phys. Rev. Mater.}, vol.~6, no.~2, p.~024803, 2022.

\bibitem{holleck1986material}
H.~Holleck, ``Material selection for hard coatings,'' {\em J. Vac. Sci. Technol.}, vol.~4, no.~6, pp.~2661--2669, 1986.

\bibitem{wang1995electrical}
C.~Wang, S.~Akbar, W.~Chen, and V.~Patton, ``Electrical properties of high-temperature oxides, borides, carbides, and nitrides,'' {\em J. Mater. Sci.}, vol.~30, pp.~1627--1641, 1995.

\bibitem{waldl2022evolution}
H.~Waldl, M.~Tkadletz, A.~Lechner, C.~Czettl, M.~Pohler, and N.~Schalk, ``Evolution of the fracture properties of arc evaporated {Ti}$_{1-x}${Al}$_{x}${N} coatings with increasing {Al} content,'' {\em Surf. Coat. Technol.}, vol.~444, p.~128690, 2022.

\bibitem{moritz2021microstructure}
Y.~Moritz, C.~Kainz, M.~Tkadletz, C.~Czettl, M.~Pohler, and N.~Schalk, ``Microstructure and mechanical properties of arc evaporated {Ti} ({Al}, {Si}) {N} coatings,'' {\em Surf. Coat. Technol.}, vol.~421, p.~127461, 2021.

\bibitem{daniel2017grain}
R.~Daniel, M.~Meindlhumer, W.~Baumegger, J.~Zalesak, B.~Sartory, M.~Burghammer, C.~Mitterer, and J.~Keckes, ``Grain boundary design of thin films: using tilted brittle interfaces for multiple crack deflection toughening,'' {\em Acta Mater.}, vol.~122, pp.~130--137, 2017.

\bibitem{csanadi2020small}
T.~Csan{\'a}di, M.~Vojtko, Z.~Dankh{\'a}zi, M.~J. Reece, and J.~Dusza, ``Small scale fracture and strength of high-entropy carbide grains during microcantilever bending experiments,'' {\em J. Eur. Ceram. Soc.}, vol.~40, no.~14, pp.~4774--4782, 2020.

\bibitem{tatami2015local}
J.~Tatami, M.~Katayama, M.~Ohnishi, T.~Yahagi, T.~Takahashi, T.~Horiuchi, M.~Yokouchi, K.~Yasuda, D.~K. Kim, T.~Wakihara, {\em et~al.}, ``Local fracture toughness of {Si}$_{3}${N}$_{4}$ ceramics measured using single-edge notched microcantilever beam specimens,'' {\em J. Am. Ceram. Soc.}, vol.~98, no.~3, pp.~965--971, 2015.

\bibitem{best2016small}
J.~P. Best, J.~Zechner, J.~M. Wheeler, R.~Schoeppner, M.~Morstein, and J.~Michler, ``Small-scale fracture toughness of ceramic thin films: the effects of specimen geometry, ion beam notching and high temperature on chromium nitride toughness evaluation,'' {\em Phil. Mag.}, vol.~96, no.~32-34, pp.~3552--3569, 2016.

\bibitem{monteverde2003advances}
F.~Monteverde, S.~Guicciardi, and A.~Bellosi, ``Advances in microstructure and mechanical properties of zirconium diboride based ceramics,'' {\em Mater. Sci. Eng. A}, vol.~346, no.~1-2, pp.~310--319, 2003.

\bibitem{csanadi2024effect}
T.~Csan{\'a}di, A.~Azizpour, M.~Vojtko, and W.~G. Fahrenholtz, ``The effect of crystal anisotropy on fracture toughness and strength of {ZrB}$_{2}$ microcantilevers,'' {\em J. Am. Ceram. Soc.}, vol.~107, no.~3, pp.~1669--1681, 2024.

\bibitem{vidivs2024hardness}
M.~Vidi{\v{s}}, T.~Fiantok, M.~Gocn{\'\i}k, P.~{\v{S}}vec~Jr, {\v{S}}.~Nagy, M.~Truchl{\`y}, V.~Izai, T.~Roch, L.~Satrapinskyy, V.~{\v{S}}roba, {\em et~al.}, ``Hardness and fracture toughness enhancement in transition metal diboride multilayer films with structural variations,'' {\em Materialia}, vol.~34, p.~102070, 2024.

\bibitem{hu2024influence}
C.~Hu, S.~Lin, M.~Podsednik, S.~Mr{\'a}z, T.~Wojcik, A.~Limbeck, N.~Koutn{\'a}, and P.~H. Mayrhofer, ``Influence of co-sputtering {AlB}$_2$ to {TaB}$_2$ on stoichiometry of non-reactively sputtered boride thin films,'' {\em Mater. Res. Lett.}, vol.~12, no.~8, pp.~561--570, 2024.

\bibitem{glechner2022influence}
T.~Glechner, H.~Oemer, T.~Wojcik, M.~Weiss, A.~Limbeck, J.~Ramm, P.~Polcik, and H.~Riedl, ``Influence of si on the oxidation behavior of tm-si-b2{\k{a}}z coatings (tm= ti, cr, hf, ta, w),'' {\em urf. Coat. Technol.}, vol.~434, p.~128178, 2022.

\bibitem{fuger2023tissue}
C.~Fuger, R.~Hahn, A.~Hirle, T.~Wojcik, P.~Kutrowatz, F.~Bohrn, O.~Hunold, P.~Polcik, and H.~Riedl, ``Tissue phase affected fracture toughness of nano-columnar {TiB}$_{2+z}$ thin films,'' {\em Mater. Res. Lett.}, vol.~11, no.~8, pp.~613--622, 2023.

\bibitem{zhou2015general}
Y.~Zhou, H.~Xiang, Z.~Feng, and Z.~Li, ``General trends in electronic structure, stability, chemical bonding and mechanical properties of ultrahigh temperature ceramics {TMB}$_{2}$ ({TM}= transition metal),'' {\em J. Mater. Sci. \& technology}, vol.~31, no.~3, pp.~285--294, 2015.

\bibitem{gan2021robust}
Q.~Gan, H.~Liu, S.~Zhang, F.~Wang, J.~Cheng, X.~Wang, S.~Dong, Q.~Tao, Y.~Chen, and P.~Zhu, ``Robust hydrophobic materials by surface modification in transition-metal diborides,'' {\em ACS Appl. Mater. Interfaces}, vol.~13, no.~48, pp.~58162--58169, 2021.

\bibitem{sun2017anisotropic}
W.~Sun, H.~Xiang, F.-Z. Dai, J.~Liu, and Y.~Zhou, ``Anisotropic surface stability of {TiB}$_{2}$: A theoretical explanation for the easy grain coarsening,'' {\em J. Mater. Res.}, vol.~32, no.~14, pp.~2755--2763, 2017.

\bibitem{sun2016theoretical}
W.~Sun, J.~Liu, H.~Xiang, and Y.~Zhou, ``A theoretical investigation on the anisotropic surface stability and oxygen adsorption behavior of {ZrB}$_{2}$,'' {\em J. Am. Ceram. Soc.}, vol.~99, no.~12, pp.~4113--4120, 2016.

\bibitem{yang2023first}
T.~Yang, X.~Han, W.~Li, X.~Chen, and P.~Liu, ``First-principles calculations on the interfacial stability and bonding properties of {HfN} (111)/{HfB}$_{2}$ (0001) interface,'' {\em Vacuum}, vol.~207, p.~111678, 2023.

\bibitem{hu2025microstructure}
C.~Hu, S.~Mr{\'a}z, P.~J. P{\"o}llmann, T.~Wojcik, M.~Podsednik, B.~Hajas, A.~Limbeck, N.~Koutn{\'a}, J.~M. Schneider, and P.~H. Mayrhofer, ``Microstructure, mechanical properties, thermal decomposition and oxidation sequences of crystalline {AlB}$_2$ thin films,'' {\em Mater. Des.}, vol.~250, p.~113584, 2025.

\bibitem{MV}
E.~V. Podryabinkin and A.~V. Shapeev, ``Active learning of linearly parametrized interatomic potentials,'' {\em Comput. Mater. Sci.}, vol.~140, pp.~171--180, 2017.

\bibitem{sangiovanni2021temperature}
D.~G. Sangiovanni, F.~Tasn{\'a}di, T.~Harrington, M.~Od{\'e}n, K.~S. Vecchio, and I.~A. Abrikosov, ``Temperature-dependent elastic properties of binary and multicomponent high-entropy refractory carbides,'' {\em Mater. Des.}, vol.~204, p.~109634, 2021.

\bibitem{sangiovanni2023valence}
D.~G. Sangiovanni, K.~Kaufmann, and K.~Vecchio, ``Valence electron concentration as key parameter to control the fracture resistance of refractory high-entropy carbides,'' {\em Sci. Adv.}, vol.~9, no.~37, p.~eadi2960, 2023.

\bibitem{koutna2022atomistic}
N.~Koutn{\'a}, L.~L{\"o}fler, D.~Holec, Z.~Chen, Z.~Zhang, L.~Hultman, P.~H. Mayrhofer, and D.~G. Sangiovanni, ``Atomistic mechanisms underlying plasticity and crack growth in ceramics: a case study of {AlN/TiN} superlattices,'' {\em Acta Mater.}, vol.~229, p.~117809, 2022.

\bibitem{Paidar2020}
V.~Paidar and J.~\v{C}apek, ``Anisotropy of fracture in hexagonal metals,'' {\em Int. J. Fract.}, vol.~225, pp.~123--127, 2020.

\bibitem{Kaushik2014}
V.~Kaushik, R.~Narasimhan, and R.~K. Mishra, ``Experimental study of fracture behavior of magnesium single crystals,'' {\em Mater. Sci. amp; Eng. A}, vol.~590, pp.~174--185, 2014.

\bibitem{Mine1998}
Y.~Mine, S.~Ando, and K.~Takashima, ``Fatigue crack propagation in titanium single crystals,'' {\em Key Eng. Mater.}, vol.~145-149, pp.~721--726, 1998.

\bibitem{andric2018atomistic}
P.~Andric and W.~A. Curtin, ``Atomistic modeling of fracture,'' {\em Model. Simul. Mat. Sci. Eng.}, vol.~27, no.~1, p.~013001, 2018.

\bibitem{sangiovanni2024controlled}
D.~G. Sangiovanni, A.~Kjell{\'e}n, F.~Trybel, L.~Johnson, M.~Od{\'e}n, F.~Tasn{\'a}di, and I.~Abrikosov, ``Controlled polymorphic competition--a path to tough and hard ceramics,'' {\em Acta Materialia}, vol.~294, p.~121121, 2025.

\bibitem{Thomson1971}
R.~Thomson, C.~Hsieh, and V.~Rana, ``Lattice trapping of fracture cracks,'' {\em J. Appl. Phys.}, vol.~42, no.~8, pp.~3154--3160, 1971.

\bibitem{andric2019atomistic}
P.~Andric and W.~A. Curtin, ``Atomistic modeling of fracture,'' {\em Model. Simul. Mater. Sci. Eng.}, vol.~27, no.~1, p.~013001, 2019.

\bibitem{sangiovanni2023descriptor}
D.~G. Sangiovanni, A.~Kraych, M.~Mrovec, J.~Salamania, M.~Od{\'e}n, F.~Tasn{\'a}di, and I.~A. Abrikosov, ``Descriptor for slip-induced crack blunting in refractory ceramics,'' {\em Phys. Rev. Mater.}, vol.~7, no.~10, p.~103601, 2023.

\bibitem{ferber1983effect}
M.~K. Ferber, P.~F. Becher, and C.~B. Finch, ``Effect of microstructure on the properties of {TiB}$_{2}$ ceramics,'' {\em J. Am. Ceram. Soc.}, vol.~66, no.~1, pp.~C--2, 1983.

\bibitem{bhaumik2000synthesis}
S.~Bhaumik, C.~Divakar, A.~K. Singh, and G.~Upadhyaya, ``Synthesis and sintering of {TiB}$_{2}$ and {TiB}$_{2}$--{TiC} composite under high pressure,'' {\em Mater. Sci. Eng. A}, vol.~279, no.~1-2, pp.~275--281, 2000.

\bibitem{wang2002influence}
W.~Wang, Z.~Fu, H.~Wang, and R.~Yuan, ``Influence of hot pressing sintering temperature and time on microstructure and mechanical properties of {TiB}$_{2}$ ceramics,'' {\em J. Eur. Ceram. Soc.}, vol.~22, no.~7, pp.~1045--1049, 2002.

\bibitem{swab2023mechanical}
J.~J. Swab, J.~Jarman, W.~Fahrenholtz, and J.~Watts, ``Mechanical properties of {ZrB}$_{2}$ ceramics determined by two laboratories,'' {\em Int. J. Appl. Ceram. Technol.}, vol.~20, no.~5, pp.~3097--3103, 2023.

\bibitem{li2024synthesis}
K.~Li, Z.~Huang, J.~Yuan, X.~Li, Z.~Wang, M.~Hu, T.~Wang, X.~Hu, Y.~Li, and X.~Zhang, ``Synthesis and growth mechanism of highly crystalized multi-branched {HfB}$_{2}$ microrods with self-toughening effect,'' {\em Mater. Des.}, vol.~244, p.~113196, 2024.

\bibitem{wang2015fabrication}
Z.~Wang, X.~Liu, B.~Xu, and Z.~Wu, ``Fabrication and properties of {HfB}$_{2}$ ceramics based on micron and submicron {HfB}$_{2}$ powders synthesized via carbo/borothermal reduction of {HfO}$_{2}$ with {B}$_{4}${C} and carbon,'' {\em Int. J. Refract. Met. Hard Mater.}, vol.~51, pp.~130--136, 2015.

\bibitem{maccagno1985brittle}
T.~Maccagno and J.~Knott, ``Brittle fracture under mixed modes {I} and {II} loading,'' {\em Int. J. Fract.}, vol.~29, pp.~R49--R57, 1985.

\bibitem{royer1988study}
J.~Royer, ``Study of pure and mixed-mode fracture of a brittle material,'' {\em Exp. Mech.}, vol.~28, pp.~382--387, 1988.

\bibitem{de2009pure}
M.~De~Moura, R.~Campilho, and J.~Gon{\c{c}}alves, ``Pure mode {II} fracture characterization of composite bonded joints,'' {\em Int. J. Solids Struct.}, vol.~46, no.~6, pp.~1589--1595, 2009.

\bibitem{jalayer2023novel}
R.~Jalayer, B.~Saboori, and M.~R. Ayatollahi, ``A novel test specimen for mixed mode {I}/{II}/{III} fracture study in brittle materials,'' {\em Fatigue Fract. Eng. Mater. Struct.}, vol.~46, no.~5, pp.~1908--1920, 2023.

\bibitem{rizov2013mixed}
V.~Rizov, ``Mixed-mode {I}/{II} fracture study of polymer composites using single edge notched bend specimens,'' {\em Comput. Mater. Sci.}, vol.~77, pp.~1--6, 2013.

\bibitem{braham2010laboratory}
A.~Braham, W.~Buttlar, and F.~Ni, ``Laboratory mixed-mode cracking of asphalt concrete using the single-edge notch beam,'' {\em Road Mater. Pavement Des.}, vol.~11, no.~4, pp.~947--968, 2010.

\bibitem{margevicius1999fracture}
R.~Margevicius, J.~Riedle, and P.~Gumbsch, ``Fracture toughness of polycrystalline tungsten under mode {I} and mixed mode {I}/{II} loading,'' {\em Mater. Sci. Eng. A}, vol.~270, no.~2, pp.~197--209, 1999.

\bibitem{sih1974}
G.~Sih and B.~MacDonald, ``Fracture mechanics applied to engineering problems -- strain energy density fracture criterion,'' {\em Engineering Fracture Mechanics}, vol.~6, no.~2, pp.~361--386, 1974.

\bibitem{zou2025socket}
Y.~Zou, B.~Derreberry, and M.~Farooq, ``Through-wall failure of a main steam bypass line socket weld and the flaw evaluation methodology,'' {\em International Journal of Pressure Vessels and Piping}, vol.~216, p.~105531, 2025.

\bibitem{predan2013stress}
J.~Predan, V.~Močilnik, and N.~Gubeljak, ``Stress intensity factors for circumferential semi-elliptical surface cracks in a hollow cylinder subjected to pure torsion,'' {\em Engineering Fracture Mechanics}, vol.~105, pp.~152--168, 2013.

\bibitem{gupta2020fracture}
I.~Gupta, C.~Sondergeld, and C.~Rai, ``Fracture toughness in shales using nano-indentation,'' {\em J. Pet. Sci. Eng.}, vol.~191, p.~107222, 2020.

\bibitem{hunter2016investigations}
B.~Hunter, X.-X. Yu, N.~De~Leon, C.~Weinberger, W.~Fahrenholtz, G.~Hilmas, M.~L. Weaver, and G.~B. Thompson, ``Investigations into the slip behavior of zirconium diboride,'' {\em J. Mater. Res.}, vol.~31, no.~18, pp.~2749--2756, 2016.

\bibitem{fuger2022anisotropic}
C.~Fuger, R.~Hahn, L.~Zauner, T.~Wojcik, M.~Weiss, A.~Limbeck, O.~Hunold, P.~Polcik, and H.~Riedl, ``Anisotropic super-hardness of hexagonal {WB}$_{2\pm z}$ thin films,'' {\em Mater. Res. Lett.}, vol.~10, no.~2, pp.~70--77, 2022.

\bibitem{hirle2025mechanical}
A.~Hirle, P.~D{\"o}rflinger, C.~Fuger, C.~Gutschka, T.~Wojcik, M.~Podsednik, A.~Limbeck, S.~Kolozsv{\'a}ri, P.~Polcik, C.~Jerg, {\em et~al.}, ``Mechanical properties of dcms and hipims deposited ti1-xmoxb2$\pm$z coatings,'' {\em Surf. Coat. Technol.}, vol.~497, p.~131750, 2025.

\bibitem{palisaitis2022nature}
J.~Palisaitis, M.~Dahlqvist, L.~Hultman, I.~Petrov, J.~Rosen, and P.~O. Persson, ``On the nature of planar defects in transition metal diboride line compounds,'' {\em Materialia}, vol.~24, p.~101478, 2022.

\bibitem{palisaitis2021unpaired}
J.~Palisaitis, M.~Dahlqvist, A.~J. Hall, J.~Th{\"o}rnberg, I.~Persson, N.~Nedfors, L.~Hultman, J.~E. Greene, I.~Petrov, J.~Rosen, {\em et~al.}, ``Where is the unpaired transition metal in substoichiometric diboride line compounds?,'' {\em Acta Mater.}, vol.~204, p.~116510, 2021.

\bibitem{thornberg2020microstructure}
J.~Th{\"o}rnberg, J.~Palisaitis, N.~Hellgren, F.~F. Klimashin, N.~Ghafoor, I.~Zhirkov, C.~Azina, J.-L. Battaglia, A.~Kusiak, M.~A. Sortica, {\em et~al.}, ``Microstructure and materials properties of understoichiometric {TiB}$_x$ thin films grown by {HiPIMS},'' {\em Surf. Coat. Technol.}, vol.~404, p.~126537, 2020.

\bibitem{VASP-1}
G.~Kresse and J.~Furthm{\"u}ller, ``Efficient iterative schemes for ab initio total-energy calculations using a plane-wave basis set,'' {\em Phys. Rev. B}, vol.~54, no.~16, p.~11169, 1996.

\bibitem{VASP-2}
G.~Kresse and D.~Joubert, ``From ultrasoft pseudopotentials to the projector augmented-wave method,'' {\em Phys. Rev. B}, vol.~59, pp.~1758--1775, Jan 1999.

\bibitem{PBEsol}
J.~P. Perdew, A.~Ruzsinszky, G.~I. Csonka, O.~A. Vydrov, G.~E. Scuseria, L.~A. Constantin, X.~Zhou, and K.~Burke, ``Restoring the density-gradient expansion for exchange in solids and surfaces,'' {\em Phys. Rev. Lett.}, vol.~100, p.~136406, Apr 2008.

\bibitem{leiner2023energetics}
T.~Leiner, N.~Koutn{\'a}, J.~Janovec, M.~Zelen{\`y}, P.~H. Mayrhofer, and D.~Holec, ``On energetics of allotrope transformations in transition-metal diborides via plane-by-plane shearing,'' {\em Vacuum}, vol.~215, p.~112329, 2023.

\bibitem{NPT}
M.~Parrinello and A.~Rahman, ``Polymorphic transitions in single crystals: A new molecular dynamics method,'' {\em J. Appl. Phys.}, vol.~52, no.~12, pp.~7182--7190, 1981.

\bibitem{LAMMPS}
A.~P. Thompson, H.~M. Aktulga, R.~Berger, D.~S. Bolintineanu, W.~M. Brown, P.~S. Crozier, P.~J. in~'t Veld, A.~Kohlmeyer, S.~G. Moore, T.~D. Nguyen, R.~Shan, M.~J. Stevens, J.~Tranchida, C.~Trott, and S.~J. Plimpton, ``{LAMMPS} - a flexible simulation tool for particle-based materials modeling at the atomic, meso, and continuum scales,'' {\em Comp. Phys. Comm.}, vol.~271, p.~108171, 2022.

\bibitem{andric2017new}
P.~Andric and W.~Curtin, ``New theory for mode {I} crack-tip dislocation emission,'' {\em J. Mech. Phys. Solids}, vol.~106, pp.~315--337, 2017.

\bibitem{ting1996anisotropic}
T.~T.~C. Ting, {\em Anisotropic Elasticity: Theory and Applications}.
\newblock Oxford University Press, 04 1996.

\bibitem{schaffer2012sample}
M.~Schaffer, B.~Schaffer, and Q.~Ramasse, ``Sample preparation for atomic-resolution stem at low voltages by {FIB},'' {\em Ultramicroscopy}, vol.~114, pp.~62--71, 2012.

\end{thebibliography}
